\documentclass{article} 
\PassOptionsToPackage{table}{xcolor}
\usepackage{iclr2027_conference,times}

\usepackage{amsmath,amsfonts,bm}

\def\eqref#1{equation~\ref{#1}}

\def\1{\bm{1}}

\DeclareMathAlphabet{\mathsfit}{\encodingdefault}{\sfdefault}{m}{sl}
\SetMathAlphabet{\mathsfit}{bold}{\encodingdefault}{\sfdefault}{bx}{n}

\usepackage{hyperref}
\usepackage{url}

\title{Prompted Identity Degrades Cooperation in Multi-Agent LLM Systems}

\author{Xavier Del Giudice$^\dagger$, Alessio Palma, Matteo Migliarini, Fabio Galasso, Indro Spinelli \\
Sapienza University of Rome, Italy \\
$^\dagger$\texttt{delgiudice.1967219@studenti.uniroma1.it} \\
}

\iclrfinalcopy 

\usepackage{dsfont}
\usepackage[utf8]{inputenc} 
\usepackage[T1]{fontenc}    
\usepackage{hyperref}       
\usepackage{url}            
\usepackage{booktabs}       
\usepackage{array}
\usepackage{amsfonts}       
\usepackage{nicefrac}       
\usepackage{microtype}      
\usepackage{xcolor}         
\definecolor{amiRamp}{HTML}{D55E00}
\usepackage{amsmath}
\usepackage{graphicx}
\usepackage{tikz}
\usetikzlibrary{positioning, arrows.meta, calc, backgrounds, fit, shapes.geometric, decorations.pathreplacing, arrows.meta, positioning, calc, patterns}
\usepackage{subcaption}
\usepackage{cleveref}
\usepackage{pgfplots}
\usepackage{booktabs} 
\usepackage{multirow}
\usepackage{placeins}
\usepackage{listings}       
\definecolor{promptbg}{RGB}{247,248,250}
\definecolor{promptframe}{RGB}{215,218,224}
\lstdefinestyle{prompt}{%
  basicstyle=\ttfamily\footnotesize,
  breaklines=true,
  breakatwhitespace=true,
  columns=fullflexible,
  keepspaces=true,
  showstringspaces=false,
  frame=single,
  rulecolor=\color{promptframe},
  backgroundcolor=\color{promptbg},
  framesep=2mm,
  framerule=0.4pt,
  xleftmargin=0.01\textwidth,
  xrightmargin=0.01\textwidth,
  aboveskip=0.8em,
  belowskip=0.8em
}
\lstnewenvironment{prompt}{\lstset{style=prompt}}{}
\lstdefinestyle{dialog}{%
  basicstyle=\ttfamily\scriptsize,
  breaklines=true,
  breakatwhitespace=false,
  columns=fullflexible,
  keepspaces=true,
  showstringspaces=false,
  frame=l,
  framesep=2pt,
  framerule=0.4pt,
  xleftmargin=4pt,
  xrightmargin=0pt,
  aboveskip=2pt,
  belowskip=2pt,
  literate={\ }{{\ }}1,
  extendedchars=true,
  inputencoding=utf8
}
\pgfplotsset{compat=1.15}
\usepackage{amssymb}

\graphicspath{{images/}{images/icons/}}

\newlength{\taskh}
\expandafter\def\csname taskh@le\endcsname{0.91em}
\expandafter\def\csname taskh@ex\endcsname{1.00em}
\expandafter\def\csname taskh@gq\endcsname{0.88em}
\newcommand{\taskicon}[2][1.0]{%
  \setlength{\taskh}{\csname taskh@#2\endcsname}%
  \setlength{\taskh}{#1\taskh}%
  \raisebox{-0.20\taskh}{\includegraphics[height=\taskh]{#2}}}
\newcommand{\iconLE}[1][1.0]{\taskicon[#1]{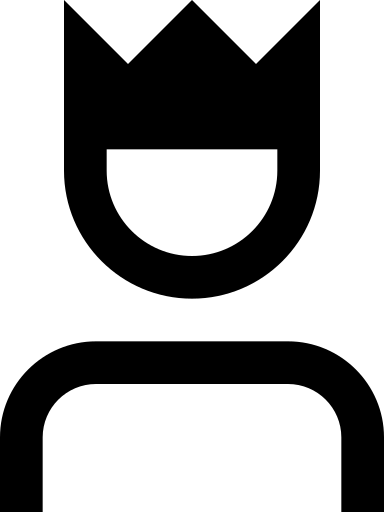}}
\newcommand{\iconEX}[1][1.0]{\taskicon[#1]{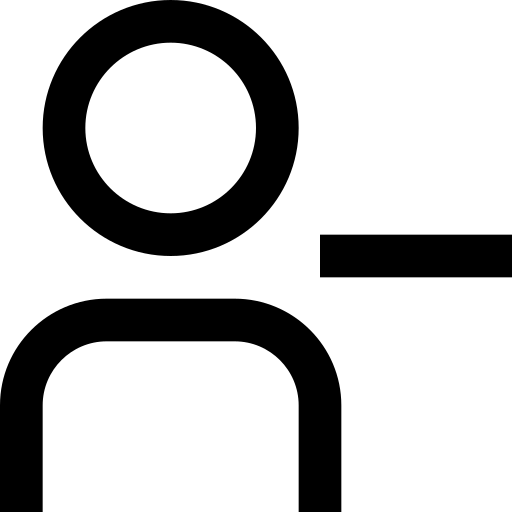}}
\newcommand{\iconGQ}[1][1.0]{\taskicon[#1]{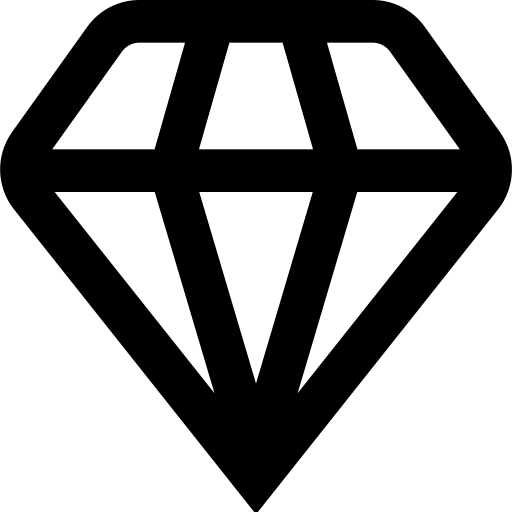}}
\newcommand{\ckpt}[1]{\texttt{\ckptscan#1\relax}}
\def\ckptscan#1{\ifx#1\relax\else#1\ifx#1-\allowbreak\fi\expandafter\ckptscan\fi}
\begin{document}

\maketitle
\lhead{Preprint} 


\begin{abstract}
Multi-agent LLM systems increasingly mix models from several providers, yet exposing each agent's underlying model identity to its peers significantly impairs cooperation. 
We show that when agents are aware of each other's model family, the group splits into clusters, where agents prefer interacting with others carrying their same label, although nothing in the task rewards or asks for such a split. 
We argue that the label itself causes this split, which we define as \textit{factionalism}. 
We show and measure this phenomenon in two cooperative games and on a reasoning benchmark, with nine to twenty-five agents drawn from up to five open-weight model families. 
We further show that when the announced families are shuffled, or replaced by arbitrary labels, the factions still follow this information; when the label is removed, this behavior disappears. 
In strictly cooperative tasks, labeled groups spend on average $30\%$ more rounds and $55\%$ more tokens to reach a decision, and their success rate drops from $96\%$ to $81\%$. The effect replicates across tasks, group sizes and model families. Withholding identity labels from the agents is simple and effective mitigation.

\begin{figure}[h]
    \centering    \includegraphics[width=\linewidth]{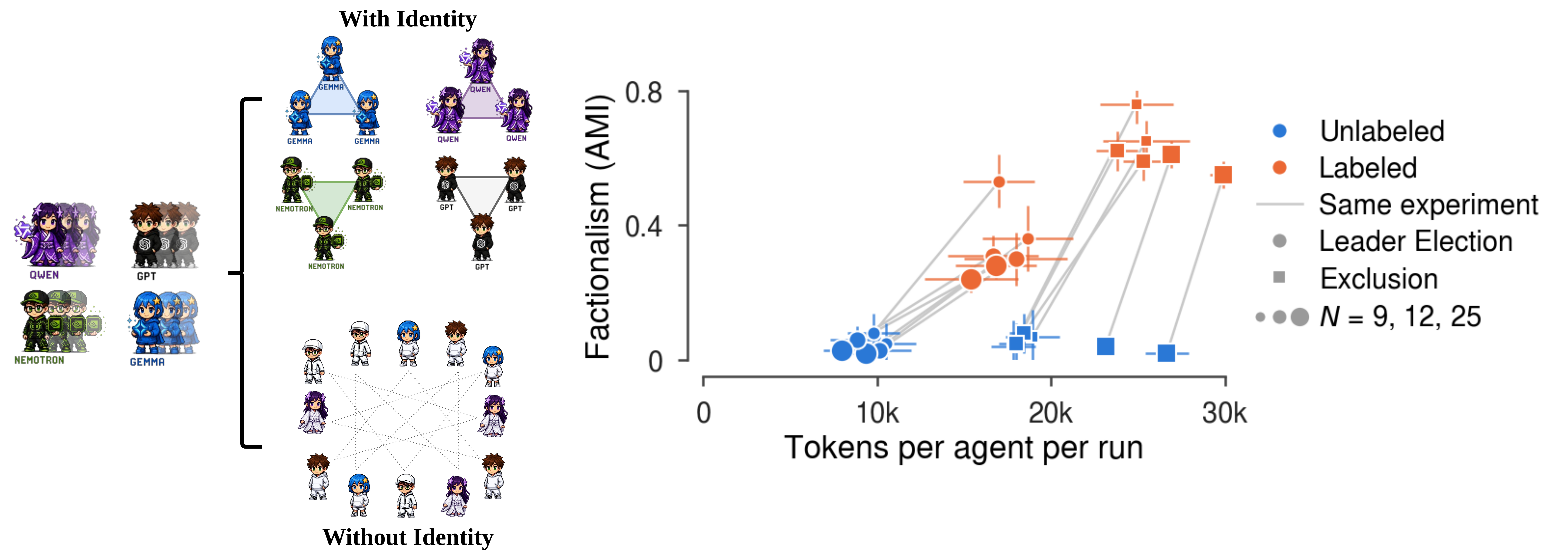}
    \caption{\textbf{Visible identity labels degrade cooperation.} In this work we place agents from several open-weight model families in cooperation games and vary only what they know about each other. When the family label is visible the group splits into label-aligned communities, and spends on average $55\%$ more tokens to reach a decision; when no identity metadata is shown, neither happens.}
    \label{fig:teaser}
\end{figure}

\end{abstract}

\section{Introduction}
\label{sec:intro}
Multi-agent systems built from large language models let several agents plan, write code, and review one another's work~\citep{wu2024autogen,hong2024metagptmetaprogrammingmultiagent,qian2024chatdevcommunicativeagentssoftware}. Large populations of agents sharing the same environment already led to unexpected results.
In a recent incident, roughly $700$ of about $1{,}200$ independently launched agents self-organized on a shared message board, with roles and task assignments that nobody designed~\citep{metr2026huggingfaceincident}. We do not yet fully understand how such systems behave. They can fail even when each constituent model acts as intended, because the failure arises from their interactions rather than from any individual model~\citep{cemri2025multiagentllmsystemsfail}.
\citet{anthropic2026multiagentpatterns} catalogs the failure modes of emerging multi-agent systems as coordination, conformity, epistemic, and goal-incompatibility failures. In this paper we address the first of these, and show that a design detail as small as what agents are told about one another changes how well a group coordinates.

Existing systems increasingly mix models from several providers, and the identity or role of each agent is a configuration detail that can reach the agents themselves, through system prompts, agent names, or message headers~\citep{wang2025mixtureofagents,chen2025internet,wu2024autogen}.
Although this information is not intended to guide their behavior, no prior work has investigated how agents act on this.
We ask whether this seemingly irrelevant piece of information changes how a heterogeneous group cooperates. 
We found that when agents are told which model family they and their peers instantiate, their interaction graph partitions along that label, although every agent is asked to pursue the same group objective and receives no persona and no explicit reward in forming groups (Figure~\ref{fig:teaser}). 
We call this emergent behaviour \emph{factionalism}, and demonstrate that it persists when family names are shuffled across agents or replaced by neutral tags such as colors, showing that it follows the visible label rather than the underlying architecture.

We study factionalism in two coordination games, \iconLE~Leader Election and \iconEX~Exclusion. 
Both games ask the agents for nothing but a common decision, a leader in one case and a peer to remove in the other. There is no correct answer, no reward for favoring a same-family peer, and no instruction to form groups, so whatever structure follows the label comes from the label itself. We run experiments with $9$ to $25$ agents from up to five open-weight model families, under three labeling conditions and a neutral-color control (Section~\ref{sec:designprinciples}), and we repeat this labeling condition design on \iconGQ~GPQA-Diamond~\citep{rein2024gpqa}. 
Factionalism appears in every experiment with a visible label and in none without.
We further show that factionalism has an immediate downstream effect: labeled groups take more rounds and more tokens to reach a decision, and succeed less often (Section~\ref{sec:results}). 
The design implication is agent-facing anonymization: agents should see only task-relevant capability descriptors, while an orchestrator retains provider identity for routing, calibration, debugging, and accountability. The contributions of our work are:
\begin{itemize}
    \item \textbf{We identify and define factionalism in heterogeneous LLM groups.} In two
    coordination games and on a question-answering benchmark, the interaction graph partitions
    following the identity labels, even though all agents are prompted to pursue a common objective and receive no incentive to form groups.

    \item \textbf{Factionalism follows visible labels, not the underlying architecture.} We design a
    \emph{mislabeled} counterfactual experiment that holds each model fixed, while shuffling the identity
    label shown to agents. The partitioning follows these shuffled tags, persists with neutral labels, and disappears when no identity metadata is shown.

    \item \textbf{Factionalism has a cost.} Announcing identities makes consensus-based tasks slower and more expensive, lowers success rates, and the cost grows with roster size.
\end{itemize}

\section{Related Work}

\noindent\textbf{Self-preference and identity cues in LLM evaluation.}
A growing body of work shows that LLMs can favor outputs associated with themselves. In single-agent evaluation settings, LLM judges exhibit self-preference, favoring their own generations over comparable alternatives \citep{zheng2023judgingllmasajudgemtbenchchatbot,xu2024prideprejudicellmamplifies,migliarini2026quantifyingselfpreservationbiaslarge}. \citet{panickssery2024llmevaluatorsrecognizefavor} link self-recognition to self-preference, while other work argues that the effect may instead arise from familiarity with low-perplexity text \citep{wataoka2025selfpreferencebiasllmasajudge}. We revisit this distinction by separating recognition of architecture-specific writing styles from the effect of prompt-visible identity labels (Section~\ref{sec:mechanism}).

Closer to our setting, \citet{choi2026identityskewsdebateanonymization} show that debate agents weight their own previous answer differently from a peer's, and reduce this bias by removing self-versus-peer attribution. Their identity cue is therefore response ownership rather than model-family membership; even their heterogeneous setting varies assigned personas within single-architecture debates. We instead retain sender attribution and manipulate the family label shown for each agent in mixed-architecture groups, asking whether categorical tags reorganize interaction into factions. Our \emph{mislabeled} and neutral-label conditions separate this effect from both architecture-specific writing style and the semantics of model names (Section~\ref{sec:designprinciples}).

\noindent\textbf{In-group bias, minimal groups, and LLM agents.}
The minimal group paradigm shows that humans can favor arbitrary in-groups even in the absence of material incentives or meaningful group differences \citep{tajfel1970experiments,tajfel1979integrative,brewer1979ingroup}. Economic models formalize this observation by incorporating group identity into utility and social preferences \citep{akerlof2000economics,chen2009group}. Recent work has begun to map these behavioral phenomena onto LLMs. Social identity biases have been documented across many language models \citep{hu2024generativelanguagemodelsexhibit}, and LLM agents can exhibit intergroup bias under explicit \textit{us-versus-them} cues \citep{wang2026agentshumansoutgroupbeliefdependent}.

Our setting differs in two respects. First, we study heterogeneous populations of interacting LLM agents rather than isolated completions or single-agent judgments. Second, the group cue is not a human social identity or an experimentally imposed team identity, but the model-family metadata commonly available in heterogeneous LLM systems. The mislabeled manipulation further distinguishes our contribution from generic in-group bias, and lets us identify prompt-visible identity metadata as the operative carrier of the effect.

\noindent\textbf{Multi-agent LLM systems and interaction structure.} Most work on LLM-based multi-agent systems evaluates aggregate task performance \citep{wu2024autogen,hong2024metagptmetaprogrammingmultiagent,qian2024chatdevcommunicativeagentssoftware,wang2024battleagentbenchbenchmarkevaluatingcooperation}, run-level failure modes \citep{cemri2025multiagentllmsystemsfail}, or coordination under imposed topologies, roles, information asymmetry, and team organization
\citep{qian2025scaling,liu2024autonomous,agashe2025llmcoordination,bhattacharyya2026social,faulkner2026elections,kumar2026cooperative}. These establish that interaction design matters, but do not ask whether incidental model metadata itself creates persistent structure in the interaction graph or affects system performance.
To our knowledge, no existing benchmark directly isolates this phenomenon. MARBLE \citep{zhu2025multiagentbenchevaluatingcollaborationcompetition} may seem close, but it assigns roles and fixes communication links between agents, while evaluating coordination only with an LLM-as-a-judge scorer.

A different line studies language-mediated coordination directly, from classical bargaining and cheap-talk theory
\citep{nash1950bargaining,rubinstein1982perfect,crawford1982strategic} to recent LLM systems that integrate dialogue with planning \citep{meta2022cicero} or target coordination as a benchmark or training objective \citep{agashe2025llmcoordination,abdelnabi2024cooperation,park2025maporl}. Our games sit in this regime of unrestricted natural-language messages under a shared team-level success condition. However, we ask a different question: not whether agents coordinate, but whether their coordination remains family-neutral internally. Work on cognitive-agent networks shows that imposed homophilic rules induce segregation-like structures \citep{Zomer2026-ja}; we instead measure factionalism that emerges from ordinary task interaction when only the family metadata visible to agents is varied.


\section{Factionalism in Multi-Agent Systems}
\label{sec:factionality}

We define \textbf{factionalism} as the tendency of agents to organize into groups that share an identity. 
Factionalism can be recognized as a partition of the interaction graph, which separates into communities aligned with identities. 
Underneath it lies \textbf{homophily}, a tendency where an agent directs more positive interaction toward peers with the same label than toward peers with a different one. 
This is the local ``birds of a feather'' mechanism~\citep{birds}, instantiated through contact or agreement between ordered agent pairs. Communities that follow the label cannot form unless agents agree more with same-label peers than with the others, so homophily is necessary but not sufficient. We therefore measure the two separately and call a group factional only when both hold (Section~\ref{sec:measurement}).

Here we study whether factionalism emerges in controlled heterogeneous LLM-agent groups. In each multi-turn game, agents from several open-weight model families pursue a shared goal and receive no individual reward. Within each run they differ only in their underlying model type, across experimental conditions we instead vary the identity labels shown to them. Section~\ref{sec:measurement} defines how to measure factionalism, which applies to any labeled multi-agent system, while Sections~\ref{sec:designprinciples} and~\ref{sec:experimentalsettings} describe the experimental design we instantiate it on.

\subsection{Measurement Framework}
\label{sec:measurement}

In every game we study, at each turn, every agent first takes one observable decision, such as casting a vote, then communicates that decision plus a short message to a few peers of its choice. We call this decision the agent's \emph{commitment}, and indicate with $v_i^{(t)}$ the commitment of agent $a_i$ at round $t$. Its \textit{recipients} $\mathcal{C}_i^{(t)} \subset \mathcal{A}_t$ are the peers it sent that message to, with $\mathcal{A}_t$ being the set of agents still active at round $t$. We infer factionalism from this trace: we convert each run into a directed weighted graph over agents, and test whether the formed communities align with the announced labels. The measurement uses only per-round commitments and recipient choices, so it can be computed for any labeled multi-agent system that logs its interaction trace.

\noindent\textbf{Graph construction.}
For a run with agents $\mathcal{A}$, we build the \emph{agreement graph} as a directed weighted graph
$\mathcal{G}=(\mathcal{A},\mathcal{E},w)$. For each ordered pair $(a_i,a_j)$, the edge weight is the rate at which $a_j$'s next-round commitment matches the commitment that $a_i$ sent when choosing $a_j$ as a recipient:
\begin{equation}
w(a_i \to a_j) =
\begin{cases}
\dfrac{\#\{t : a_j \in \mathcal{C}_i^{(t)} \cap \mathcal{A}_{t+1} \land v_j^{(t+1)} = v_i^{(t)}\}}
     {\#\{t : a_j \in \mathcal{C}_i^{(t)} \cap \mathcal{A}_{t+1}\}}
     & \text{if } \#\{t : a_j \in \mathcal{C}_i^{(t)} \cap \mathcal{A}_{t+1}\} > 0,\\[5pt]
\text{absent} & \text{otherwise.}
\end{cases}
\label{eq:edge_construction}
\end{equation}
The normalization controls for contact volume and convergence timing: a pair that remains aligned over many contacted rounds doesn't receive a greater weight just because it communicates more.
We distinguish two partitions of the agents: $\ell^{\mathrm{vis}}$ groups them by the label they see, and $\ell^{\mathrm{true}}$ by their underlying model architecture. Unlabeled runs, which have no visible labels, are evaluated against $\ell^{\mathrm{true}}$.

\noindent\textbf{Homophily.}
For the tested partition, we split ordered pairs into within-label and between-label pairs and take the ratio of their mean edge weights $w$ from Equation~\ref{eq:edge_construction}, which we call the \emph{Edge Density Ratio} (EDR). Its significance is tested through a run-specific permutation null test that reshuffles labels, while preserving tested-group sizes (details in Appendix~\ref{app:inference}). EDR measures homophily, the same-label preference without which a group cannot be factional.

\noindent\textbf{Community structure.}
For community detection, we use the same contacted agent pairs but weight each edge by the count of next-round commitment matches (the numerator of Equation~\ref{eq:edge_construction}), without normalizing by contact count. The graph is symmetrized by adding the two directions of each pair, a missing direction counting as zero. We apply Louvain community detection~\citep{Blondel_2008} to this graph and use Adjusted Mutual Information (AMI)~\citep{JMLR:v11:vinh10a} to score the alignment between the detected communities and the tested partition. Full edge-case conventions, including zero-denominator handling and permutation $p$-value correction, are given in Appendix~\ref{app:inference}.

\noindent\textbf{Decision rule.}
For each experiment, we aggregate across runs with different seeds and apply two EDR and two AMI tests (Appendix~\ref{app:inference}), then correct the four $p$-values together with Holm--Bonferroni at $\alpha=0.05$, and report \emph{factionalism} only when all four pass. In every significant factional result, EDR points in the same direction as AMI, so we report EDR values in the appendix tables only (Appendix~\ref{sec:appendix}).

\subsection{Design principles}
\label{sec:designprinciples}
Our design isolates the identity label from everything else that could align agents along family lines. A run succeeds only if the system meets the game's objective within the round limit, otherwise the whole group fails. No player gains from supporting a same-family peer or loses from being selected, excluded, or represented by another lineage. Agents receive no personas, positions, or private goals; we call games with these properties \emph{no-stakes}. The task instructions never provide explicit instrumental incentives to coordinate by tag, so any such structure arises from the models' behavior under the exposed identity metadata.
We also vary how many agents each family contributes, a roster is either \emph{balanced}, with equal family sizes, or \emph{unbalanced}, with one family over-represented: $3/3/3$ versus $4/3/2$ at $N=9$, $3/3/3/3$ versus $4/3/3/2$ at $N=12$, and $5/5/5/5/5$ versus $9/6/4/3/3$ at $N=25$. Balanced rosters rule out a numerical majority as the source of same-family preference, while unbalanced rosters test whether the effect persists despite one.

In our experiments, the manipulated variable is the family metadata exposed to the agents. Under \emph{unlabeled}, no model-family metadata is shown. Under \emph{labeled}, each agent sees its own family and the true families of its peers. Under \emph{mislabeled}, visible tags are shuffled relative to the models that actually run, so the announced identity and the underlying architecture belong to different families. Unless stated otherwise, we test the partition the agents saw. Because model-family names may carry reputational or semantic associations behind them, we repeat the \emph{labeled} and \emph{mislabeled} conditions with neutral color tokens in their place.

\subsection{Experimental settings}
\label{sec:experimentalsettings}

We instantiate the system with up to five open-weight model families: Gemma (\ckpt{gemma-4-26B-A4B-it}), GPT-OSS (\ckpt{gpt-oss-20b}), Nemotron (\ckpt{NVIDIA-Nemotron-3-Nano-30B-A3B-BF16}), Qwen (\ckpt{Qwen3.6-35B-A3B}), and GLM (\ckpt{GLM-4.7-Flash}). Model-specific sampling parameters are listed in Appendix~\ref{app:llm_models}. Experiments with rosters of $N \in \{9,12,25\}$ agents draw on three, four, and five of these families respectively, in the balanced and unbalanced compositions of Section~\ref{sec:designprinciples}.

Both no-stakes games run on the same messaging framework. At round $t$, each active agent $a_i \in \mathcal{A}_t$ emits a structured action block containing a natural-language message $m_i^{(t)}$, the game-specific commitment $v_i^{(t)}$, and exactly $k=2$ active peers as recipients $\mathcal{C}_i^{(t)}$; on the next round it observes only the payloads addressed to it. A run ends when the game's success condition is met or at the round limit $T$. Across the two games, we set $T=20$ in every setting except Exclusion at $N=25$, where $T=30$ because a win requires between $20$ and $22$ removals. Since recipient selection is part of the agent policy, family-aligned contact and agreement can emerge without being specified by the environment. Appendix~\ref{app:dialogs} gives representative trajectory excerpts.

\begin{figure}[t]
    \centering
    \includegraphics[width=\linewidth]{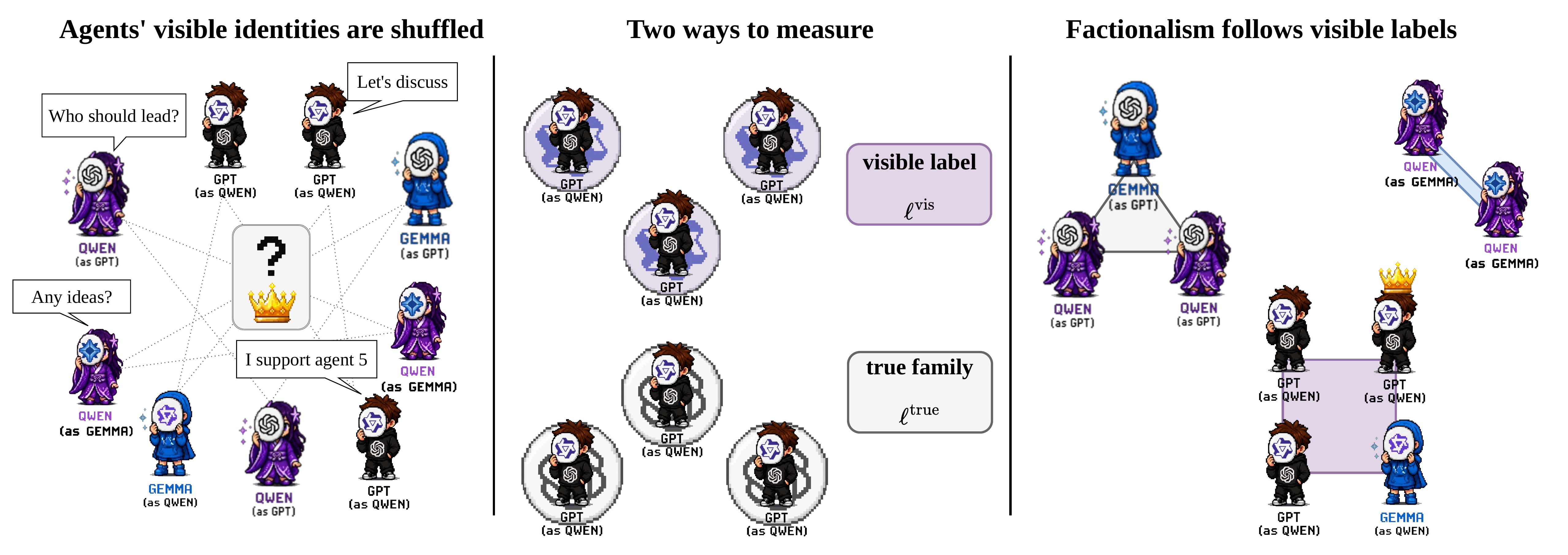}
    \caption{\textbf{The counterfactual manipulation.} \emph{Left:} under \emph{mislabeled}, every agent is told to belong to a model family which is not its actual one. \emph{Center:} the same agreement graph is evaluated both against the visible-label partition $\ell^{\mathrm{vis}}$ and the true-family partition $\ell^{\mathrm{true}}$.
    \emph{Right:} factions form around the announced tags.}
    \label{fig:games-overview}
\end{figure}

\noindent\iconLE~\textbf{Leader Election.}
Each round, every agent casts one ballot $v_i^{(t)} \in \mathcal{A}$. A run succeeds when a single candidate is the unique top vote-getter with at least $\lceil 0.75N\rceil$ ballots, otherwise it fails at horizon $T$. Figure~\ref{fig:games-overview} visualizes one Leader Election turn under the mislabeled condition.

\noindent\iconEX~\textbf{Exclusion.}
Each round, every alive agent $a_i \in \mathcal{A}_t$ casts one ballot
$v_i^{(t)} \in \mathcal{A}_t \setminus \{a_i\}$ naming a peer to remove. A unique peer receiving at least $\lfloor 2|\mathcal{A}_t|/3 \rfloor$ votes is removed; ties and sub-threshold pluralities produce no elimination. A run succeeds when $|\mathcal{A}_t| \leq k_{\mathrm{win}}$, where $k_{\mathrm{win}}$ is the smallest initial family size, so that any family could in principle supply the surviving cohort.

\section{Results}
\label{sec:results}

We report two main findings after executing $50$ runs for each reported configuration. \textbf{(i)} Visible family labels induce factionalism, while anonymization suppresses it: all $24$ labeled or mislabeled experiments show it (Table~\ref{tab:factionalism}) and none of the $12$ unlabeled ones does. \textbf{(ii)} Visible labels degrade outcomes: announcing families makes games slower, more expensive and lowers success. The ablations in Section~\ref{sec:results-ablations} test alternative explanations for factionalism; Section~\ref{sec:results-mechanism} examines how visible labels affect whose decisions agents adopt.

\subsection{Visible labels induce factionalism}
\label{sec:results-factionality}

Table~\ref{tab:factionalism} reports results for every roster size and setting. AMI is near zero when the communities found in the agreement graph are unrelated to the analyzed labels, and one when the two coincide.

All unlabeled experiments stay near the permutation baseline, with AMI between $0.02$ and $0.08$: removing agent-facing family labels suppresses factionalism. When true labels are shown, all experiments pass the decision rule. Factionalism is as strong on balanced rosters as on unbalanced ones, so a numerical majority is not the main driver. In Exclusion, AMI lies between $0.55$ and $0.76$, while in Leader Election the range is $[0.24, 0.53]$.
Even in the mislabeled setting, all the results are significant, with AMI between $0.18$ and $0.35$ in Leader Election and between $0.55$ and $0.69$ in Exclusion. The system therefore splits along the announced identity, although that identity doesn't match the real architecture. 
In the analyzed runs, message content is consistent with coordination around visible labels. Agents name model families only when labels are shown, and coordination terms become more frequent under shuffled than truthful labels (Appendix~\ref{app:messages}).

\begin{table}[htbp]
\centering
\caption{AMI between communities in the agreement graph and the tested partition: true families for \emph{unlabeled} runs, visible labels otherwise. Subscripts are standard errors, darker is more factional and significant results are in \textbf{bold}.}
\label{tab:factionalism}
\setlength{\tabcolsep}{4pt}
\renewcommand{\arraystretch}{1.15}
\small
\begin{tabular}{llcccccc}
\toprule
\multirow{2}{*}{\textbf{Composition}} & \multirow{2}{*}{\textbf{Condition}} & \multicolumn{3}{c}{\iconLE[1.25]\,\textbf{Leader Election}} & \multicolumn{3}{c}{\iconEX[1.25]\,\textbf{Exclusion}} \\
\cmidrule(lr){3-5}\cmidrule(lr){6-8}
 & & $N=9$ & $N=12$ & $N=25$ & $N=9$ & $N=12$ & $N=25$ \\
\midrule
\multirow{3}{*}{Unbalanced} & Unlabeled & \cellcolor{amiRamp!3}$0.05_{\pm 0.03}$ & \cellcolor{amiRamp!2}$0.03_{\pm 0.02}$ & \cellcolor{amiRamp!1}$0.02_{\pm 0.01}$ & \cellcolor{amiRamp!5}$0.07_{\pm 0.04}$ & \cellcolor{amiRamp!5}$0.08_{\pm 0.03}$ & \cellcolor{amiRamp!1}$0.02_{\pm 0.01}$ \\
 & Labeled & \cellcolor{amiRamp!30}$\mathbf{0.43}_{\pm 0.05}$ & \cellcolor{amiRamp!21}$\mathbf{0.30}_{\pm 0.04}$ & \cellcolor{amiRamp!16}$\mathbf{0.24}_{\pm 0.02}$ & \cellcolor{amiRamp!45}$\mathbf{0.65}_{\pm 0.03}$ & \cellcolor{amiRamp!41}$\mathbf{0.59}_{\pm 0.03}$ & \cellcolor{amiRamp!38}$\mathbf{0.55}_{\pm 0.02}$ \\
 & Mislabeled & \cellcolor{amiRamp!23}$\mathbf{0.33}_{\pm 0.05}$ & \cellcolor{amiRamp!12}$\mathbf{0.18}_{\pm 0.03}$ & \cellcolor{amiRamp!14}$\mathbf{0.21}_{\pm 0.02}$ & \cellcolor{amiRamp!46}$\mathbf{0.67}_{\pm 0.03}$ & \cellcolor{amiRamp!40}$\mathbf{0.58}_{\pm 0.03}$ & \cellcolor{amiRamp!38}$\mathbf{0.55}_{\pm 0.03}$ \\
\midrule
\multirow{3}{*}{Balanced} & Unlabeled & \cellcolor{amiRamp!5}$0.08_{\pm 0.03}$ & \cellcolor{amiRamp!4}$0.06_{\pm 0.02}$ & \cellcolor{amiRamp!2}$0.03_{\pm 0.01}$ & \cellcolor{amiRamp!3}$0.04_{\pm 0.04}$ & \cellcolor{amiRamp!3}$0.05_{\pm 0.03}$ & \cellcolor{amiRamp!3}$0.04_{\pm 0.01}$ \\
 & Labeled & \cellcolor{amiRamp!36}$\mathbf{0.53}_{\pm 0.04}$ & \cellcolor{amiRamp!21}$\mathbf{0.31}_{\pm 0.03}$ & \cellcolor{amiRamp!19}$\mathbf{0.28}_{\pm 0.02}$ & \cellcolor{amiRamp!52}$\mathbf{0.76}_{\pm 0.03}$ & \cellcolor{amiRamp!43}$\mathbf{0.62}_{\pm 0.03}$ & \cellcolor{amiRamp!42}$\mathbf{0.61}_{\pm 0.02}$ \\
 & Mislabeled & \cellcolor{amiRamp!24}$\mathbf{0.35}_{\pm 0.04}$ & \cellcolor{amiRamp!16}$\mathbf{0.23}_{\pm 0.03}$ & \cellcolor{amiRamp!16}$\mathbf{0.24}_{\pm 0.02}$ & \cellcolor{amiRamp!47}$\mathbf{0.69}_{\pm 0.04}$ & \cellcolor{amiRamp!40}$\mathbf{0.58}_{\pm 0.03}$ & \cellcolor{amiRamp!41}$\mathbf{0.59}_{\pm 0.02}$ \\
\bottomrule
\end{tabular}
\end{table}

\subsection{The cost of visible labels}
\label{sec:results-cost}

We record whether the group succeeds, the rounds it spends, and the tokens agents generate in messages and reasoning. For each task and roster size, we fit one generalized linear model per outcome~\citep{mccullagh1989generalized}, comparing \emph{labeled} and \emph{mislabeled} runs with \emph{unlabeled} ones. Each model includes a covariate for roster balance, so the reported label effects are adjusted for that difference. We report ratios rather than absolute differences: $1.4$ means $40\%$ more rounds or tokens, while $0.8$ for success means a $20\%$ lower success rate. 
Table~\ref{tab:cost} reports the \emph{labeled} comparisons, and more details can be found in Appendix~\ref{app:cost}.

Announcing families slows both scenarios, makes them more expensive, and lowers success rate. Leader Election takes $1.4\times$ rounds to converge, and generates $1.8\times$ tokens across every roster size. Exclusion runs are already close to the $T$-round limit even without labels, so part of the slowdown can't show up in the count: rounds anyway rise by $9$--$16\%$, while tokens rise by $1.4\times$ at $N=9$ and $N=12$. Success also falls more as the roster grows: the labeled-to-unlabeled success-probability ratio drops from $0.96$ to $0.86$ in Leader Election and from $0.89$ to $0.61$ in Exclusion between $N=9$ and $N=12$. Because labeled groups both spend more tokens per run and succeed less often, the tokens paid per successful run rise by a factor between $1.5$ and $2.2$, increasing with roster size in both games (Appendix~\ref{app:cost}). At $N=25$ no labeled or mislabeled Exclusion run converges within the horizon while only $8$ of $100$ unlabeled runs do so. Therefore we choose not to report them. 

\begin{table}[htbp]
\centering
\caption{Effects of truthful labels relative to \emph{unlabeled}, adjusted for roster balance. Entries are ratios, with standard errors as subscripts and significant effects in \textbf{bold}; dashes denote unavailable estimates.}
\label{tab:cost}
\setlength{\tabcolsep}{5pt}
\small
\begin{tabular}{llccc}
\toprule
\textbf{Task} & $\mathbf{N}$ & \textbf{Rounds spent} & \textbf{Generated tokens} & \textbf{Success} \\
\midrule
\multirow{3}{*}{\textbf{Leader Election}} & $9$ & $\mathbf{1.42}_{\pm 0.09}$ & $\mathbf{1.76}_{\pm 0.13}$ & $\mathbf{0.96}_{\pm 0.02}$ \\
 & $12$ & $\mathbf{1.46}_{\pm 0.10}$ & $\mathbf{1.82}_{\pm 0.16}$ & $\mathbf{0.86}_{\pm 0.04}$ \\
 & $25$ & $\mathbf{1.38}_{\pm 0.09}$ & $\mathbf{1.86}_{\pm 0.14}$ & $\mathbf{0.85}_{\pm 0.04}$ \\
\midrule
\multirow{3}{*}{\textbf{Exclusion}} & $9$ & $\mathbf{1.16}_{\pm 0.04}$ & $\mathbf{1.37}_{\pm 0.06}$ & $\mathbf{0.89}_{\pm 0.04}$ \\
 & $12$ & $\mathbf{1.09}_{\pm 0.02}$ & $\mathbf{1.35}_{\pm 0.03}$ & $\mathbf{0.61}_{\pm 0.08}$ \\
 & $25$ & -- & $\mathbf{1.14}_{\pm 0.02}$ & -- \\
\midrule
\multirow{2}{*}{\textbf{GPQA-Diamond}} & $9$ & $\mathbf{1.31}_{\pm 0.05}$ & $\mathbf{1.10}_{\pm 0.03}$ & $\mathbf{0.91}_{\pm 0.03}$ \\
 & $12$ & $\mathbf{1.15}_{\pm 0.03}$ & $\mathbf{1.09}_{\pm 0.03}$ & $\mathbf{0.89}_{\pm 0.04}$ \\
\bottomrule
\end{tabular}
\end{table}

\subsection{Ablation studies}
\label{sec:results-ablations}

Three controlled variations test whether the results in Table~\ref{tab:factionalism} depend on factors other than visible labels: shuffled labels analysis separates tags from model architecture recognition, neutral colors remove family-name associations, and GPQA-Diamond introduces an independently specified correct answer.

\noindent\textbf{Factionalism follows visible labels.}
Under \emph{mislabeled}, each agent's announced family differs from its model family. We detect communities once from each run's agreement graph, then compare them with the visible-label partition ($\ell^{\mathrm{vis}}$; Table~\ref{tab:factionalism}) and the true-family partition ($\ell^{\mathrm{true}}$; Table~\ref{tab:true-partition}). If architecture were the main driver, alignment with true families would remain strong. Instead, the visible-label partition meets the factionalism criterion in every experiment, while the true-family partition does so only in the two unbalanced $N=9$ experiments. These exceptions are explained by the assignment constraints: with a $4/3/2$ roster, preserving label counts while giving every agent a false label forces all members of one true family to share a visible label, creating overlap between the partitions. The grouping, therefore, follows what agents are told more closely than the architecture of the models generating the messages, as confirmed also by the stylistic probe in Section~\ref{sec:results-mechanism}.
\begin{table}[htbp]
\centering
\caption{AMI between communities in the agreement graph and true model families in \emph{mislabeled} runs. Table~\ref{tab:factionalism} reports visible-label AMI for the same runs. Subscripts are standard errors; \textbf{bold} marks experiments meeting the factionalism criterion.}
\label{tab:true-partition}
\setlength{\tabcolsep}{4pt}
\renewcommand{\arraystretch}{1.15}
\small
\begin{tabular}{lcccccc}
\toprule
\multirow{2}{*}{\textbf{Composition}} & \multicolumn{3}{c}{\iconLE[1.25]\,\textbf{Leader Election}} & \multicolumn{3}{c}{\iconEX[1.25]\,\textbf{Exclusion}} \\
\cmidrule(lr){2-4}\cmidrule(lr){5-7}
 & $N=9$ & $N=12$ & $N=25$ & $N=9$ & $N=12$ & $N=25$ \\
\midrule
Unbalanced & \cellcolor{amiRamp!12}$\mathbf{0.18}_{\pm 0.03}$ & \cellcolor{amiRamp!3}$0.04_{\pm 0.03}$ & \cellcolor{amiRamp!5}$0.08_{\pm 0.01}$ & \cellcolor{amiRamp!20}$\mathbf{0.29}_{\pm 0.03}$ & \cellcolor{amiRamp!8}$0.11_{\pm 0.03}$ & \cellcolor{amiRamp!5}$0.08_{\pm 0.01}$ \\
Balanced & \cellcolor{amiRamp!7}$0.10_{\pm 0.03}$ & \cellcolor{amiRamp!3}$0.04_{\pm 0.02}$ & \cellcolor{amiRamp!5}$0.07_{\pm 0.02}$ & \cellcolor{amiRamp!9}$0.13_{\pm 0.03}$ & \cellcolor{amiRamp!2}$0.03_{\pm 0.02}$ & \cellcolor{amiRamp!3}$0.05_{\pm 0.01}$ \\
\bottomrule
\end{tabular}
\end{table}

\noindent\textbf{Neutral labels.}
Because model family names carry prior associations of reputation and provenance, the observed effect could stem from these preconceptions rather than emergent categorical distinction alone. We then repeat the \emph{labeled} and \emph{mislabeled} conditions with one neutral token per family, drawn from five colors (\textit{teal}, \textit{coral}, \textit{silver}, \textit{violet}, \textit{maroon}). All color-labeled experiments satisfy the factionalism criterion, and under color mislabeling, segregation consistently aligns with visible color partitions rather than true underlying architectures (Table~\ref{tab:colors}). Consequently, factionalism emerges even under arbitrary labels devoid of reputational cues.

\begin{table}[htbp]
\centering
\caption{AMI with neutral color labels. Unlabeled baselines as in Table~\ref{tab:factionalism}. Subscripts are standard errors; significant results in \textbf{bold}.}
\label{tab:colors}
\setlength{\tabcolsep}{2.5pt}
\renewcommand{\arraystretch}{1.15}
\small
\begin{tabular}{llcccccc}
\toprule
\multirow{2}{*}{\textbf{Composition}} & \multirow{2}{*}{\textbf{Condition}} & \multicolumn{3}{c}{\iconLE[1.25]\,\textbf{Leader Election}} & \multicolumn{3}{c}{\iconEX[1.25]\,\textbf{Exclusion}} \\
\cmidrule(lr){3-5}\cmidrule(lr){6-8}
 & & $N=9$ & $N=12$ & $N=25$ & $N=9$ & $N=12$ & $N=25$ \\
\midrule
\multirow{3}{*}{Unbalanced} & Labeled & \cellcolor{amiRamp!25}$\mathbf{0.37}_{\pm 0.05}$ & \cellcolor{amiRamp!14}$\mathbf{0.21}_{\pm 0.03}$ & \cellcolor{amiRamp!16}$\mathbf{0.24}_{\pm 0.02}$ & \cellcolor{amiRamp!45}$\mathbf{0.66}_{\pm 0.03}$ & \cellcolor{amiRamp!47}$\mathbf{0.69}_{\pm 0.03}$ & \cellcolor{amiRamp!55}$\mathbf{0.80}_{\pm 0.02}$ \\
 & Mislabeled $\ell^{\mathrm{vis}}$ & \cellcolor{amiRamp!21}$\mathbf{0.31}_{\pm 0.04}$ & \cellcolor{amiRamp!13}$\mathbf{0.19}_{\pm 0.03}$ & \cellcolor{amiRamp!12}$\mathbf{0.18}_{\pm 0.02}$ & \cellcolor{amiRamp!52}$\mathbf{0.76}_{\pm 0.02}$ & \cellcolor{amiRamp!49}$\mathbf{0.72}_{\pm 0.03}$ & \cellcolor{amiRamp!54}$\mathbf{0.79}_{\pm 0.02}$ \\
 & Mislabeled $\ell^{\mathrm{true}}$ & \cellcolor{amiRamp!1}$0.01_{\pm 0.03}$ & \cellcolor{amiRamp!1}$0.02_{\pm 0.02}$ & \cellcolor{amiRamp!5}$0.07_{\pm 0.01}$ & \cellcolor{amiRamp!0}$-0.12_{\pm 0.02}$ & \cellcolor{amiRamp!0}$-0.02_{\pm 0.02}$ & \cellcolor{amiRamp!0}$-0.01_{\pm 0.01}$ \\
\midrule
\multirow{3}{*}{Balanced} & Labeled & \cellcolor{amiRamp!30}$\mathbf{0.43}_{\pm 0.04}$ & \cellcolor{amiRamp!17}$\mathbf{0.25}_{\pm 0.03}$ & \cellcolor{amiRamp!18}$\mathbf{0.26}_{\pm 0.02}$ & \cellcolor{amiRamp!55}$\mathbf{0.82}_{\pm 0.03}$ & \cellcolor{amiRamp!46}$\mathbf{0.67}_{\pm 0.03}$ & \cellcolor{amiRamp!55}$\mathbf{0.81}_{\pm 0.02}$ \\
 & Mislabeled $\ell^{\mathrm{vis}}$ & \cellcolor{amiRamp!19}$\mathbf{0.28}_{\pm 0.04}$ & \cellcolor{amiRamp!14}$\mathbf{0.21}_{\pm 0.03}$ & \cellcolor{amiRamp!14}$\mathbf{0.21}_{\pm 0.02}$ & \cellcolor{amiRamp!55}$\mathbf{0.95}_{\pm 0.02}$ & \cellcolor{amiRamp!52}$\mathbf{0.75}_{\pm 0.03}$ & \cellcolor{amiRamp!55}$\mathbf{0.84}_{\pm 0.02}$ \\
 & Mislabeled $\ell^{\mathrm{true}}$ & \cellcolor{amiRamp!0}$-0.03_{\pm 0.03}$ & \cellcolor{amiRamp!4}$0.06_{\pm 0.02}$ & \cellcolor{amiRamp!4}$0.06_{\pm 0.01}$ & \cellcolor{amiRamp!0}$-0.05_{\pm 0.02}$ & \cellcolor{amiRamp!0}$-0.01_{\pm 0.02}$ & \cellcolor{amiRamp!1}$0.01_{\pm 0.01}$ \\
\bottomrule
\end{tabular}
\end{table}

\noindent\iconGQ~\textbf{GPQA-Diamond.}
To test whether factionalism persists with objective ground truth answers rather than self-referential consensus, we replicate our Leader Election protocol on GPQA-Diamond (Appendix~\ref{app:gpqa}).
Agents answer each question while exchanging fan-out-$2$ messages until a leader is elected or the $T$ rounds terminate. The answer of the elected leader is submitted as the group's answer, and a run succeeds if that answer is correct.

Table~\ref{tab:gpqa} reveals the same pattern: unlabeled runs show no clustering, identity labels induce factionalism, shuffled labels steer alignment toward the announced tags and only very weakly to the true families.
This happens even though correctness is determined by an external objective answer and there is no reward for following peers.
We report identity-produced prices in Table~\ref{tab:cost}: labeled groups take $1.31\times$ and $1.15\times$ rounds at $N=9$ and $N=12$, generate $1.10\times$ and $1.09\times$ tokens at $N=9$ and $N=12$, and answer correctly less often, with success ratios of $0.91\times$ and $0.89\times$.


\begin{table}[htbp]
\centering
\caption{AMI on \iconGQ~GPQA-Diamond. Subscripts are standard errors; significant results in \textbf{bold}.}
\label{tab:gpqa}
\setlength{\tabcolsep}{4pt}
\renewcommand{\arraystretch}{1.15}
\small
\begin{tabular}{llcc}
\toprule
\multirow{2}{*}{\textbf{Composition}} & \multirow{2}{*}{\textbf{Condition}} & \multicolumn{2}{c}{\iconGQ[1.25]\,\textbf{GPQA-Diamond}} \\
\cmidrule(lr){3-4}
 & & $N=9$ & $N=12$ \\
\midrule
\multirow{4}{*}{Unbalanced} & Unlabeled & \cellcolor{amiRamp!3}$0.05_{\pm 0.02}$ & \cellcolor{amiRamp!0}$0.00_{\pm 0.01}$ \\
 & Labeled & \cellcolor{amiRamp!19}$\mathbf{0.28}_{\pm 0.02}$ & \cellcolor{amiRamp!22}$\mathbf{0.32}_{\pm 0.02}$ \\
 & Mislabeled $\ell^{\mathrm{vis}}$ & \cellcolor{amiRamp!19}$\mathbf{0.27}_{\pm 0.02}$ & \cellcolor{amiRamp!12}$\mathbf{0.17}_{\pm 0.02}$ \\
 & Mislabeled $\ell^{\mathrm{true}}$ & \cellcolor{amiRamp!11}$\mathbf{0.17}_{\pm 0.02}$ & \cellcolor{amiRamp!4}$0.07_{\pm 0.01}$ \\
\midrule
\multirow{4}{*}{Balanced} & Unlabeled & \cellcolor{amiRamp!1}$0.02_{\pm 0.01}$ & \cellcolor{amiRamp!1}$0.02_{\pm 0.01}$ \\
 & Labeled & \cellcolor{amiRamp!27}$\mathbf{0.40}_{\pm 0.02}$ & \cellcolor{amiRamp!24}$\mathbf{0.35}_{\pm 0.02}$ \\
 & Mislabeled $\ell^{\mathrm{vis}}$ & \cellcolor{amiRamp!20}$\mathbf{0.30}_{\pm 0.02}$ & \cellcolor{amiRamp!22}$\mathbf{0.31}_{\pm 0.01}$ \\
 & Mislabeled $\ell^{\mathrm{true}}$ & \cellcolor{amiRamp!10}$0.14_{\pm 0.02}$ & \cellcolor{amiRamp!8}$0.13_{\pm 0.01}$ \\
\bottomrule
\end{tabular}
\end{table}

\subsection{How the label acts}
\label{sec:results-mechanism}
\label{sec:mechanism}

In order to further probe how identity labels shape agents' decisions, we test two channels by scoring log-probabilities under each agent's true model. For the \emph{stylistic channel} we score raw peer messages, this signal should track true architecture and remain invariant to visible labels~\citep{wataoka2025selfpreferencebiasllmasajudge}. For the \emph{label channel}, we score fixed peer-vote strings in each agent's pre-response context from actual runs. If visible labels guide adoption, this gap should be small under \emph{unlabeled}, grow under \emph{labeled}, and follow assigned rather than true families under \emph{mislabeled}.

\noindent\textbf{Stylistic score.}
For each ordered pair $(i,j)$, we score $j$'s raw message $m_j^{(r)}$ in each round $r$ under $i$'s model:
\begin{equation}
s^{\mathrm{style}}_{i \to j} =
\frac{\sum_r \log P_i(m_j^{(r)})}
{\sum_r |m_j^{(r)}|_{\mathrm{char}}}.
\label{eq:ppl_matrix}
\end{equation}
We normalize by character count to avoid tokenizer effects. Within each trajectory, we center an evaluator's scores across peers, compute its mean score for same-group peers minus its mean for different-group peers, and average these gaps with equal weight across the evaluators' underlying model families (details in Appendix~\ref{app:mechanism}). Figure~\ref{fig:e1-within-between} shows that this gap is small (at most $0.07$ nats/char) and it is present under \emph{unlabeled}, \emph{labeled}, and follows true architecture rather than the announced tag in \emph{mislabeled}. Thus, detectable stylistic similarity cannot explain the much larger label-aligned shifts in factionalism.

\noindent\textbf{Adoption score.}
For each ordered pair $(i,j)$ and round $r$, we score the fixed vote string \texttt{VOTE: $v_j^{(r)}$} under $i$'s model, conditioned on $i$'s pre-response context $c_i^{(r)}$ from the run:
\begin{equation}
s^{\mathrm{adoption}}_{i \to j,r}
=
\overline{\log P}_{i}\!\left(\texttt{VOTE: }v_j^{(r)} \mid c_i^{(r)}\right),
\label{eq:adoption_probe}
\end{equation}
where $v_j^{(r)}$ is $j$'s vote and $\overline{\log P}$ is log-probability per token. Higher scores mean that $i$ assigns more probability to that vote. We keep the round-level scores as separate observations and apply the same centered, equal-family within-minus-between comparison described for the stylistic score. Figure~\ref{fig:e4-gap} shows very small gaps under \emph{unlabeled} and larger gaps under \emph{labeled}, while in \emph{mislabeled} it follows the announced label rather than true architecture. Unlike the stylistic score, adoption thus aligns with the visible groups. Per-family results are in Tables~\ref{tab:mechanism-e4-le} and~\ref{tab:mechanism-e4-ex}.


\definecolor{condUnlabeled}{HTML}{6E6E6E}
\definecolor{condLabeled}{HTML}{0072B2}
\definecolor{condMislabeled}{HTML}{D55E00}
\definecolor{axisGray}{HTML}{555555}
\definecolor{gridGray}{HTML}{D8D8D8}

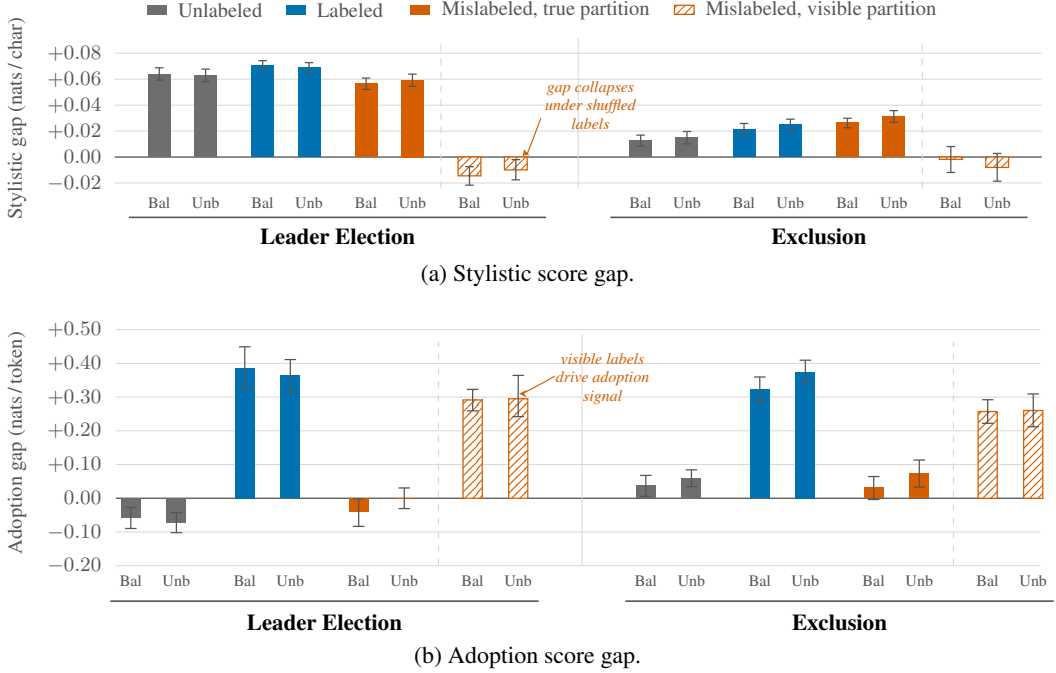
\begin{figure*}[t]
\centering

\begin{subfigure}[t]{\textwidth}
\centering
\resizebox{\linewidth}{!}{%
\begin{tikzpicture}[
  xscale=0.55,
  yscale=20,
  every node/.style={font=\footnotesize},
]

\useasboundingbox (-3.617, -0.069) rectangle (26.0, 0.120);

\foreach \y/\lab in {%
  -0.02/$-0.02$,
   0.00/$\phantom{-}0.00$,
   0.02/$+0.02$,
   0.04/$+0.04$,
   0.06/$+0.06$,
   0.08/$+0.08$%
}{%
  \draw[gridGray, line width=0.3pt] (-0.4, \y) -- (25.6, \y);
  \node[anchor=east, text=axisGray] at (-0.55, \y) {\lab};
}

\node[rotate=90, anchor=north, text=axisGray, font=\small]
  at (-3.617, 0.030) {Stylistic gap (nats\,/\,char)};

\draw[axisGray, line width=0.55pt] (-0.4, 0.0) -- (25.6, 0.0);


\fill[condUnlabeled] (0.53, 0) rectangle (1.17, 0.0639);
\draw[axisGray, line width=0.55pt] (0.85, 0.0593) -- (0.85, 0.0688);
\draw[axisGray, line width=0.55pt] (0.73, 0.0593) -- (0.97, 0.0593);
\draw[axisGray, line width=0.55pt] (0.73, 0.0688) -- (0.97, 0.0688);

\fill[condUnlabeled] (1.83, 0) rectangle (2.47, 0.0630);
\draw[axisGray, line width=0.55pt] (2.15, 0.0581) -- (2.15, 0.0678);
\draw[axisGray, line width=0.55pt] (2.03, 0.0581) -- (2.27, 0.0581);
\draw[axisGray, line width=0.55pt] (2.03, 0.0678) -- (2.27, 0.0678);

\fill[condLabeled] (3.43, 0) rectangle (4.07, 0.0703);
\draw[axisGray, line width=0.55pt] (3.75, 0.0663) -- (3.75, 0.0743);
\draw[axisGray, line width=0.55pt] (3.63, 0.0663) -- (3.87, 0.0663);
\draw[axisGray, line width=0.55pt] (3.63, 0.0743) -- (3.87, 0.0743);

\fill[condLabeled] (4.73, 0) rectangle (5.37, 0.0689);
\draw[axisGray, line width=0.55pt] (5.05, 0.0649) -- (5.05, 0.0728);
\draw[axisGray, line width=0.55pt] (4.93, 0.0649) -- (5.17, 0.0649);
\draw[axisGray, line width=0.55pt] (4.93, 0.0728) -- (5.17, 0.0728);

\fill[condMislabeled] (6.33, 0) rectangle (6.97, 0.0566);
\draw[axisGray, line width=0.55pt] (6.65, 0.0521) -- (6.65, 0.0609);
\draw[axisGray, line width=0.55pt] (6.53, 0.0521) -- (6.77, 0.0521);
\draw[axisGray, line width=0.55pt] (6.53, 0.0609) -- (6.77, 0.0609);

\fill[condMislabeled] (7.63, 0) rectangle (8.27, 0.0594);
\draw[axisGray, line width=0.55pt] (7.95, 0.0545) -- (7.95, 0.0639);
\draw[axisGray, line width=0.55pt] (7.83, 0.0545) -- (8.07, 0.0545);
\draw[axisGray, line width=0.55pt] (7.83, 0.0639) -- (8.07, 0.0639);

\draw[gridGray, line width=0.5pt, dashed] (8.75, -0.024) -- (8.75, 0.087);

\fill[white] (9.23, 0) rectangle (9.87, -0.0145);
\fill[pattern=north east lines, pattern color=condMislabeled]
  (9.23, 0) rectangle (9.87, -0.0145);
\draw[condMislabeled, line width=0.55pt] (9.23, 0) rectangle (9.87, -0.0145);
\draw[axisGray, line width=0.55pt] (9.55, -0.0074) -- (9.55, -0.0217);
\draw[axisGray, line width=0.55pt] (9.43, -0.0074) -- (9.67, -0.0074);
\draw[axisGray, line width=0.55pt] (9.43, -0.0217) -- (9.67, -0.0217);

\fill[white] (10.53, 0) rectangle (11.17, -0.0099);
\fill[pattern=north east lines, pattern color=condMislabeled]
  (10.53, 0) rectangle (11.17, -0.0099);
\draw[condMislabeled, line width=0.55pt] (10.53, 0) rectangle (11.17, -0.0099);
\draw[axisGray, line width=0.55pt] (10.85, -0.0020) -- (10.85, -0.0176);
\draw[axisGray, line width=0.55pt] (10.73, -0.0020) -- (10.97, -0.0020);
\draw[axisGray, line width=0.55pt] (10.73, -0.0176) -- (10.97, -0.0176);

\draw[gridGray, line width=0.4pt] (12.60, -0.024) -- (12.60, 0.092);

\fill[condUnlabeled] (14.03, 0) rectangle (14.67, 0.0127);
\draw[axisGray, line width=0.55pt] (14.35, 0.0084) -- (14.35, 0.0169);
\draw[axisGray, line width=0.55pt] (14.23, 0.0084) -- (14.47, 0.0084);
\draw[axisGray, line width=0.55pt] (14.23, 0.0169) -- (14.47, 0.0169);

\fill[condUnlabeled] (15.33, 0) rectangle (15.97, 0.0148);
\draw[axisGray, line width=0.55pt] (15.65, 0.0102) -- (15.65, 0.0197);
\draw[axisGray, line width=0.55pt] (15.53, 0.0102) -- (15.77, 0.0102);
\draw[axisGray, line width=0.55pt] (15.53, 0.0197) -- (15.77, 0.0197);

\fill[condLabeled] (16.93, 0) rectangle (17.57, 0.0213);
\draw[axisGray, line width=0.55pt] (17.25, 0.0165) -- (17.25, 0.0259);
\draw[axisGray, line width=0.55pt] (17.13, 0.0165) -- (17.37, 0.0165);
\draw[axisGray, line width=0.55pt] (17.13, 0.0259) -- (17.37, 0.0259);

\fill[condLabeled] (18.23, 0) rectangle (18.87, 0.0248);
\draw[axisGray, line width=0.55pt] (18.55, 0.0205) -- (18.55, 0.0292);
\draw[axisGray, line width=0.55pt] (18.43, 0.0205) -- (18.67, 0.0205);
\draw[axisGray, line width=0.55pt] (18.43, 0.0292) -- (18.67, 0.0292);

\fill[condMislabeled] (19.83, 0) rectangle (20.47, 0.0263);
\draw[axisGray, line width=0.55pt] (20.15, 0.0226) -- (20.15, 0.0299);
\draw[axisGray, line width=0.55pt] (20.03, 0.0226) -- (20.27, 0.0226);
\draw[axisGray, line width=0.55pt] (20.03, 0.0299) -- (20.27, 0.0299);

\fill[condMislabeled] (21.13, 0) rectangle (21.77, 0.0312);
\draw[axisGray, line width=0.55pt] (21.45, 0.0267) -- (21.45, 0.0358);
\draw[axisGray, line width=0.55pt] (21.33, 0.0267) -- (21.57, 0.0267);
\draw[axisGray, line width=0.55pt] (21.33, 0.0358) -- (21.57, 0.0358);

\draw[gridGray, line width=0.5pt, dashed] (22.25, -0.024) -- (22.25, 0.087);

\fill[white] (22.73, 0) rectangle (23.37, -0.0018);
\fill[pattern=north east lines, pattern color=condMislabeled]
  (22.73, 0) rectangle (23.37, -0.0018);
\draw[condMislabeled, line width=0.55pt] (22.73, 0) rectangle (23.37, -0.0018);
\draw[axisGray, line width=0.55pt] (23.05, 0.0081) -- (23.05, -0.0119);
\draw[axisGray, line width=0.55pt] (22.93, 0.0081) -- (23.17, 0.0081);
\draw[axisGray, line width=0.55pt] (22.93, -0.0119) -- (23.17, -0.0119);

\fill[white] (24.03, 0) rectangle (24.67, -0.0080);
\fill[pattern=north east lines, pattern color=condMislabeled]
  (24.03, 0) rectangle (24.67, -0.0080);
\draw[condMislabeled, line width=0.55pt] (24.03, 0) rectangle (24.67, -0.0080);
\draw[axisGray, line width=0.55pt] (24.35, 0.0028) -- (24.35, -0.0186);
\draw[axisGray, line width=0.55pt] (24.23, 0.0028) -- (24.47, 0.0028);
\draw[axisGray, line width=0.55pt] (24.23, -0.0186) -- (24.47, -0.0186);

\foreach \x in {0.85, 3.75, 6.65, 9.55, 14.35, 17.25, 20.15, 23.05}{
  \node[anchor=north, text=axisGray, font=\scriptsize] at (\x, -0.0245) {Bal};
}
\foreach \x in {2.15, 5.05, 7.95, 10.85, 15.65, 18.55, 21.45, 24.35}{
  \node[anchor=north, text=axisGray, font=\scriptsize] at (\x, -0.0245) {Unb};
}

\draw[axisGray, line width=0.7pt] (0.00, -0.0465) -- (11.70, -0.0465);
\draw[axisGray, line width=0.7pt] (13.50, -0.0465) -- (25.20, -0.0465);
\node[anchor=north, font=\bfseries] at (5.85, -0.0485) {Leader Election};
\node[anchor=north, font=\bfseries] at (19.35, -0.0485) {Exclusion};

\node[anchor=west, font=\scriptsize\itshape, text=condMislabeled, align=center]
  at (11.45, 0.040) {gap collapses\\under shuffled\\labels};
\draw[condMislabeled, ->, line width=0.5pt, >=stealth]
  (11.55, 0.03) .. controls (11.25, 0.008) and (11.15, -0.001) .. (11.08, 0.006);

\fill[condUnlabeled] (0.50, 0.110) rectangle (1.05, 0.117);
\node[anchor=west, text=axisGray] at (1.15, 0.1135) {Unlabeled};

\fill[condLabeled] (4.40, 0.110) rectangle (4.95, 0.117);
\node[anchor=west, text=axisGray] at (5.05, 0.1135) {Labeled};

\fill[condMislabeled] (7.90, 0.110) rectangle (8.45, 0.117);
\node[anchor=west, text=axisGray] at (8.55, 0.1135) {Mislabeled, true partition};

\fill[white] (15.30, 0.110) rectangle (15.85, 0.133);
\fill[pattern=north east lines, pattern color=condMislabeled]
  (15.30, 0.110) rectangle (15.85, 0.117);
\draw[condMislabeled, line width=0.55pt] (15.30, 0.110) rectangle (15.85, 0.117);
\node[anchor=west, text=axisGray] at (15.95, 0.1135) {Mislabeled, visible partition};

\end{tikzpicture}%
}
\caption{Stylistic score gap.}
\label{fig:e1-within-between}
\end{subfigure}

\vspace{0.6em}

\begin{subfigure}[t]{\textwidth}
\centering
\resizebox{\linewidth}{!}{%
\begin{tikzpicture}[
  xscale=0.703,
  yscale=5.2,
  every node/.style={font=\footnotesize},
]

\useasboundingbox (-2.870, -0.385) rectangle (20.30, 0.565);

\foreach \y/\lab in {%
  -0.20/$-0.20$,
  -0.10/$-0.10$,
   0.00/$\phantom{-}0.00$,
   0.10/$+0.10$,
   0.20/$+0.20$,
   0.30/$+0.30$,
   0.40/$+0.40$,
   0.50/$+0.50$%
}{%
  \draw[gridGray, line width=0.3pt] (-0.33, \y) -- (19.95, \y);
  \node[anchor=east, text=axisGray] at (-0.47, \y) {\lab};
}

\node[rotate=90, anchor=north, text=axisGray, font=\small]
  at (-2.870, 0.165) {Adoption gap (nats\,/\,token)};

\draw[axisGray, line width=0.55pt] (-0.33, 0.0) -- (19.95, 0.0);



\fill[condUnlabeled] (-0.2100, 0) rectangle (0.2100, -0.0591);
\draw[axisGray, line width=0.55pt] (0.0000, -0.0896) -- (0.0000, -0.0272);
\draw[axisGray, line width=0.55pt] (-0.1260, -0.0896) -- (0.1260, -0.0896);
\draw[axisGray, line width=0.55pt] (-0.1260, -0.0272) -- (0.1260, -0.0272);

\fill[condUnlabeled] (0.7900, 0) rectangle (1.2100, -0.0719);
\draw[axisGray, line width=0.55pt] (1.0000, -0.1020) -- (1.0000, -0.0424);
\draw[axisGray, line width=0.55pt] (0.8740, -0.1020) -- (1.1260, -0.1020);
\draw[axisGray, line width=0.55pt] (0.8740, -0.0424) -- (1.1260, -0.0424);

\fill[condLabeled] (2.2900, 0) rectangle (2.7100, 0.3869);
\draw[axisGray, line width=0.55pt] (2.5000, 0.3253) -- (2.5000, 0.4490);
\draw[axisGray, line width=0.55pt] (2.3740, 0.3253) -- (2.6260, 0.3253);
\draw[axisGray, line width=0.55pt] (2.3740, 0.4490) -- (2.6260, 0.4490);

\fill[condLabeled] (3.2900, 0) rectangle (3.7100, 0.3633);
\draw[axisGray, line width=0.55pt] (3.5000, 0.3147) -- (3.5000, 0.4112);
\draw[axisGray, line width=0.55pt] (3.3740, 0.3147) -- (3.6260, 0.3147);
\draw[axisGray, line width=0.55pt] (3.3740, 0.4112) -- (3.6260, 0.4112);

\fill[condMislabeled] (4.7900, 0) rectangle (5.2100, -0.0411);
\draw[axisGray, line width=0.55pt] (5.0000, -0.0834) -- (5.0000, -0.0013);
\draw[axisGray, line width=0.55pt] (4.8740, -0.0834) -- (5.1260, -0.0834);
\draw[axisGray, line width=0.55pt] (4.8740, -0.0013) -- (5.1260, -0.0013);

\fill[condMislabeled] (5.7900, 0) rectangle (6.2100, -0.0005);
\draw[axisGray, line width=0.55pt] (6.0000, -0.0306) -- (6.0000, 0.0306);
\draw[axisGray, line width=0.55pt] (5.8740, -0.0306) -- (6.1260, -0.0306);
\draw[axisGray, line width=0.55pt] (5.8740, 0.0306) -- (6.1260, 0.0306);

\draw[gridGray, line width=0.5pt, dashed] (6.7500, -0.200) -- (6.7500, 0.465);

\fill[white] (7.2900, 0) rectangle (7.7100, 0.2915);
\fill[pattern=north east lines, pattern color=condMislabeled]
  (7.2900, 0) rectangle (7.7100, 0.2915);
\draw[condMislabeled, line width=0.55pt] (7.2900, 0) rectangle (7.7100, 0.2915);
\draw[axisGray, line width=0.55pt] (7.5000, 0.2592) -- (7.5000, 0.3228);
\draw[axisGray, line width=0.55pt] (7.3740, 0.2592) -- (7.6260, 0.2592);
\draw[axisGray, line width=0.55pt] (7.3740, 0.3228) -- (7.6260, 0.3228);

\fill[white] (8.2900, 0) rectangle (8.7100, 0.2949);
\fill[pattern=north east lines, pattern color=condMislabeled]
  (8.2900, 0) rectangle (8.7100, 0.2949);
\draw[condMislabeled, line width=0.55pt] (8.2900, 0) rectangle (8.7100, 0.2949);
\draw[axisGray, line width=0.55pt] (8.5000, 0.2420) -- (8.5000, 0.3642);
\draw[axisGray, line width=0.55pt] (8.3740, 0.2420) -- (8.6260, 0.2420);
\draw[axisGray, line width=0.55pt] (8.3740, 0.3642) -- (8.6260, 0.3642);

\draw[gridGray, line width=0.4pt] (9.9000, -0.200) -- (9.9000, 0.485);


\fill[condUnlabeled] (11.0900, 0) rectangle (11.5100, 0.0375);
\draw[axisGray, line width=0.55pt] (11.3000, 0.0058) -- (11.3000, 0.0677);
\draw[axisGray, line width=0.55pt] (11.1740, 0.0058) -- (11.4260, 0.0058);
\draw[axisGray, line width=0.55pt] (11.1740, 0.0677) -- (11.4260, 0.0677);

\fill[condUnlabeled] (12.0900, 0) rectangle (12.5100, 0.0595);
\draw[axisGray, line width=0.55pt] (12.3000, 0.0345) -- (12.3000, 0.0843);
\draw[axisGray, line width=0.55pt] (12.1740, 0.0345) -- (12.4260, 0.0345);
\draw[axisGray, line width=0.55pt] (12.1740, 0.0843) -- (12.4260, 0.0843);

\fill[condLabeled] (13.5900, 0) rectangle (14.0100, 0.3227);
\draw[axisGray, line width=0.55pt] (13.8000, 0.2875) -- (13.8000, 0.3596);
\draw[axisGray, line width=0.55pt] (13.6740, 0.2875) -- (13.9260, 0.2875);
\draw[axisGray, line width=0.55pt] (13.6740, 0.3596) -- (13.9260, 0.3596);

\fill[condLabeled] (14.5900, 0) rectangle (15.0100, 0.3744);
\draw[axisGray, line width=0.55pt] (14.8000, 0.3396) -- (14.8000, 0.4093);
\draw[axisGray, line width=0.55pt] (14.6740, 0.3396) -- (14.9260, 0.3396);
\draw[axisGray, line width=0.55pt] (14.6740, 0.4093) -- (14.9260, 0.4093);

\fill[condMislabeled] (16.0900, 0) rectangle (16.5100, 0.0313);
\draw[axisGray, line width=0.55pt] (16.3000, -0.0038) -- (16.3000, 0.0644);
\draw[axisGray, line width=0.55pt] (16.1740, -0.0038) -- (16.4260, -0.0038);
\draw[axisGray, line width=0.55pt] (16.1740, 0.0644) -- (16.4260, 0.0644);

\fill[condMislabeled] (17.0900, 0) rectangle (17.5100, 0.0739);
\draw[axisGray, line width=0.55pt] (17.3000, 0.0329) -- (17.3000, 0.1133);
\draw[axisGray, line width=0.55pt] (17.1740, 0.0329) -- (17.4260, 0.0329);
\draw[axisGray, line width=0.55pt] (17.1740, 0.1133) -- (17.4260, 0.1133);

\draw[gridGray, line width=0.5pt, dashed] (18.0500, -0.200) -- (18.0500, 0.465);

\fill[white] (18.5900, 0) rectangle (19.0100, 0.2570);
\fill[pattern=north east lines, pattern color=condMislabeled]
  (18.5900, 0) rectangle (19.0100, 0.2570);
\draw[condMislabeled, line width=0.55pt] (18.5900, 0) rectangle (19.0100, 0.2570);
\draw[axisGray, line width=0.55pt] (18.8000, 0.2223) -- (18.8000, 0.2921);
\draw[axisGray, line width=0.55pt] (18.6740, 0.2223) -- (18.9260, 0.2223);
\draw[axisGray, line width=0.55pt] (18.6740, 0.2921) -- (18.9260, 0.2921);

\fill[white] (19.5900, 0) rectangle (20.0100, 0.2602);
\fill[pattern=north east lines, pattern color=condMislabeled]
  (19.5900, 0) rectangle (20.0100, 0.2602);
\draw[condMislabeled, line width=0.55pt] (19.5900, 0) rectangle (20.0100, 0.2602);
\draw[axisGray, line width=0.55pt] (19.8000, 0.2120) -- (19.8000, 0.3092);
\draw[axisGray, line width=0.55pt] (19.6740, 0.2120) -- (19.9260, 0.2120);
\draw[axisGray, line width=0.55pt] (19.6740, 0.3092) -- (19.9260, 0.3092);

\foreach \x in {0.0000, 2.5000, 5.0000, 7.5000, 11.3000, 13.8000, 16.3000, 18.8000}{
  \node[anchor=north, text=axisGray, font=\scriptsize] at (\x, -0.205) {Bal};
}
\foreach \x in {1.0000, 3.5000, 6.0000, 8.5000, 12.3000, 14.8000, 17.3000, 19.8000}{
  \node[anchor=north, text=axisGray, font=\scriptsize] at (\x, -0.205) {Unb};
}

\draw[axisGray, line width=0.7pt] (-0.45, -0.305) -- (8.95, -0.305);
\draw[axisGray, line width=0.7pt] (10.85, -0.305) -- (20.25, -0.305);
\node[anchor=north, font=\bfseries] at (4.25, -0.320) {Leader Election};
\node[anchor=north, font=\bfseries] at (15.55, -0.320) {Exclusion};

\node[anchor=west, font=\scriptsize\itshape, text=condMislabeled, align=center]
  at (9.15, 0.355) {visible labels\\drive adoption\\ signal};
\draw[condMislabeled, ->, line width=0.5pt, >=stealth]
  (9.30, 0.362) .. controls (8.70, 0.315) and (8.55, 0.310) .. (8.52, 0.310);

\end{tikzpicture}%
}
\caption{Adoption score gap.}
\label{fig:e4-gap}
\end{subfigure}

\caption{\textbf{Stylistic and adoption gaps.}
(a) Same-group minus different-group gap in the per-character stylistic score $s^{\mathrm{style}}_{i\to j}$ of raw peer messages.
(b) The corresponding gap in the per-token adoption score $s^{\mathrm{adoption}}_{i\to j,r}$ of peers' realized votes, scored in the receiving agent's pre-response context.
Both panels show balanced (Bal) and unbalanced (Unb) $N=12$ rosters; error bars are $95\%$ bootstrap confidence intervals. Groups are true model families under \emph{unlabeled} and \emph{labeled}; under \emph{mislabeled}, solid bars use true families and hatched bars use visible labels.}
\label{fig:recognition-adoption-gaps}
\end{figure*}


\section{Limitations}
\label{sec:limitations}
Our claims are deliberately scoped. We study up to five open-weight model families under controlled text-only consensus games, with a four-family check under matched sampling parameters; bigger roster sizes, closed models, other decoding regimes, and richer agentic workflows remain untested. Given that no established benchmark satisfies our no-stakes and evaluation requirements, we designed the coordination games for this study. GPQA-Diamond also demonstrates label-induced coordination costs on a benchmark with ground truth answers, but it uses the Leader Election interaction protocol. Whether these costs generalize to other externally evaluated tasks remains open.

\section{Conclusion}
Visible identity labels can induce factional structure in heterogeneous multi-agent systems, even when the task provides no incentive to form groups. The effect follows the announced identity rather than the underlying architecture, extends to neutral shared tags, and always impairs coordination. A practical implication is to avoid exposing unnecessary shared identifiers to agents: provider and model metadata can remain available to an orchestrator for routing and auditing, without entering the agents' interaction context. More broadly, our results suggest that cooperation in heterogeneous agent systems is partly an architectural choice: by controlling which identities agents can observe, we can design multi-agent systems that preserve diversity without turning it into division.

\bibliography{biblio}

@inproceedings{hong2024metagptmetaprogrammingmultiagent,
      title={Meta{GPT}: Meta Programming for A Multi-Agent Collaborative Framework},
      author={Sirui Hong and Mingchen Zhuge and Jonathan Chen and Xiawu Zheng and Yuheng Cheng and Jinlin Wang and Ceyao Zhang and Zili Wang and Steven Ka Shing Yau and Zijuan Lin and Liyang Zhou and Chenyu Ran and Lingfeng Xiao and Chenglin Wu and J{\"u}rgen Schmidhuber},
      booktitle={The Twelfth International Conference on Learning Representations},
      year={2024},
      url={https://openreview.net/forum?id=VtmBAGCN7o}
}

@inproceedings{qian2024chatdevcommunicativeagentssoftware,
    title = "{C}hat{D}ev: Communicative Agents for Software Development",
    author = "Qian, Chen  and
      Liu, Wei  and
      Liu, Hongzhang  and
      Chen, Nuo  and
      Dang, Yufan  and
      Li, Jiahao  and
      Yang, Cheng  and
      Chen, Weize  and
      Su, Yusheng  and
      Cong, Xin  and
      Xu, Juyuan  and
      Li, Dahai  and
      Liu, Zhiyuan  and
      Sun, Maosong",
    editor = "Ku, Lun-Wei  and
      Martins, Andre  and
      Srikumar, Vivek",
    booktitle = "Proceedings of the 62nd Annual Meeting of the Association for Computational Linguistics (Volume 1: Long Papers)",
    month = aug,
    year = "2024",
    address = "Bangkok, Thailand",
    publisher = "Association for Computational Linguistics",
    url = "https://aclanthology.org/2024.acl-long.810/",
    doi = "10.18653/v1/2024.acl-long.810",
    pages = "15174--15186"
}

@inproceedings{panickssery2024llmevaluatorsrecognizefavor,
author = {Panickssery, Arjun and Bowman, Samuel R. and Feng, Shi},
title = {LLM evaluators recognize and favor their own generations},
year = {2024},
isbn = {9798331314385},
publisher = {Curran Associates Inc.},
address = {Red Hook, NY, USA},
booktitle = {Proceedings of the 38th International Conference on Neural Information Processing Systems},
articleno = {2197},
numpages = {31},
location = {Vancouver, BC, Canada},
series = {NIPS '24}
}

@misc{wataoka2025selfpreferencebiasllmasajudge,
      title={Self-Preference Bias in {LLM}-as-a-Judge}, 
      author={Koki Wataoka and Tsubasa Takahashi and Ryokan Ri},
      year={2025},
      eprint={2410.21819},
      archivePrefix={arXiv},
      primaryClass={cs.CL},
      url={https://arxiv.org/abs/2410.21819}, 
}

@article{Blondel_2008,
doi = {10.1088/1742-5468/2008/10/P10008},
url = {https://doi.org/10.1088/1742-5468/2008/10/P10008},
year = {2008},
month = {oct},
publisher = {},
volume = {2008},
number = {10},
pages = {P10008},
author = {Blondel, Vincent D and Guillaume, Jean-Loup and Lambiotte, Renaud and Lefebvre, Etienne},
title = {Fast unfolding of communities in large networks},
journal = {Journal of Statistical Mechanics: Theory and Experiment}
}

@article{JMLR:v11:vinh10a,
  author  = {Nguyen Xuan Vinh and Julien Epps and James Bailey},
  title   = {Information Theoretic Measures for Clusterings Comparison: Variants, Properties, Normalization and Correction for Chance},
  journal = {Journal of Machine Learning Research},
  year    = {2010},
  volume  = {11},
  number  = {95},
  pages   = {2837--2854},
  url     = {http://jmlr.org/papers/v11/vinh10a.html}
}

@inproceedings{bhattacharyya2026social,
title={Social Agents: Collective Intelligence Improves {LLM} Predictions},
author={Aanisha Bhattacharyya and Abhilekh Borah and Yaman Kumar Singla and Rajiv Ratn Shah and Changyou Chen and Balaji Krishnamurthy},
booktitle={The Fourteenth International Conference on Learning Representations},
year={2026},
url={https://openreview.net/forum?id=73J3hsato3}
}

@inproceedings{qian2025scaling,
    title={Scaling Large Language Model-based Multi-Agent Collaboration},
    author={Chen Qian and Zihao Xie and YiFei Wang and Wei Liu and Kunlun Zhu and Hanchen Xia and Yufan Dang and Zhuoyun Du and Weize Chen and Cheng Yang and Zhiyuan Liu and Maosong Sun},
    booktitle={The Thirteenth International Conference on Learning Representations},
    year={2025},
    url={https://openreview.net/forum?id=K3n5jPkrU6}
}

@inproceedings{liu2024autonomous,
author = {Liu, Wei and Wang, Chenxi and Wang, Yifei and Xie, Zihao and Qiu, Rennai and Dang, Yufan and Du, Zhuoyun and Chen, Weize and Yang, Cheng and Qian, Chen},
title = {Autonomous agents for collaborative task under information asymmetry},
year = {2024},
isbn = {9798331314385},
publisher = {Curran Associates Inc.},
address = {Red Hook, NY, USA},
booktitle = {Proceedings of the 38th International Conference on Neural Information Processing Systems},
articleno = {90},
numpages = {32},
location = {Vancouver, BC, Canada},
series = {NIPS '24}
}

@inproceedings{agashe2025llmcoordination,
    title = "{LLM}-Coordination: Evaluating and Analyzing Multi-agent Coordination Abilities in Large Language Models",
    author = "Agashe, Saaket  and
      Fan, Yue  and
      Reyna, Anthony  and
      Wang, Xin Eric",
    editor = "Chiruzzo, Luis  and
      Ritter, Alan  and
      Wang, Lu",
    booktitle = "Findings of the Association for Computational Linguistics: NAACL 2025",
    month = apr,
    year = "2025",
    address = "Albuquerque, New Mexico",
    publisher = "Association for Computational Linguistics",
    url = "https://aclanthology.org/2025.findings-naacl.448/",
    doi = "10.18653/v1/2025.findings-naacl.448",
    pages = "8053--8072",
    ISBN = "979-8-89176-195-7",
}

@book{mccullagh1989generalized,
  title={Generalized Linear Models, Second Edition},
  author={McCullagh, P. and Nelder, J.A.},
  isbn={9780412317606},
  lccn={99013896},
  series={Chapman \& Hall/CRC Monographs on Statistics \& Applied Probability},
  url={https://books.google.it/books?id=h9kFH2_FfBkC},
  year={1989},
  publisher={Taylor \& Francis}
}

@article{holm1979simple,
 ISSN = {03036898, 14679469},
 URL = {http://www.jstor.org/stable/4615733},
 author = {Sture Holm},
 journal = {Scandinavian Journal of Statistics},
 number = {2},
 pages = {65--70},
 publisher = {[Board of the Foundation of the Scandinavian Journal of Statistics, Wiley]},
 title = {A Simple Sequentially Rejective Multiple Test Procedure},
 urldate = {2026-05-06},
 volume = {6},
 year = {1979}
}

@article{Fisher_1925, 
    title={Theory of Statistical Estimation}, 
    volume={22}, 
    DOI={10.1017/S0305004100009580}, 
    number={5},
    journal={Mathematical Proceedings of the Cambridge Philosophical Society},
    author={Fisher, R. A.}, 
    year={1925},
    pages={700–725}
}

@inproceedings{zheng2023judgingllmasajudgemtbenchchatbot,
title={Judging {LLM}-as-a-Judge with {MT}-Bench and Chatbot Arena},
author={Lianmin Zheng and Wei-Lin Chiang and Ying Sheng and Siyuan Zhuang and Zhanghao Wu and Yonghao Zhuang and Zi Lin and Zhuohan Li and Dacheng Li and Eric Xing and Hao Zhang and Joseph E. Gonzalez and Ion Stoica},
booktitle={Thirty-seventh Conference on Neural Information Processing Systems Datasets and Benchmarks Track},
year={2023},
url={https://openreview.net/forum?id=uccHPGDlao}
}

@inproceedings{xu2024prideprejudicellmamplifies,
    title = "Pride and Prejudice: {LLM} Amplifies Self-Bias in Self-Refinement",
    author = "Xu, Wenda and Zhu, Guanglei and Zhao, Xuandong and Pan, Liangming and Li, Lei and Wang, William",
    editor = "Ku, Lun-Wei  and
      Martins, Andre  and
      Srikumar, Vivek",
    booktitle = "Proceedings of the 62nd Annual Meeting of the Association for Computational Linguistics (Volume 1: Long Papers)",
    month = aug,
    year = "2024",
    address = "Bangkok, Thailand",
    publisher = "Association for Computational Linguistics",
    url = "https://aclanthology.org/2024.acl-long.826/",
    doi = "10.18653/v1/2024.acl-long.826",
    pages = "15474--15492"
}

@Article{hu2024generativelanguagemodelsexhibit,
author={Hu, Tiancheng and Kyrychenko, Yara and Rathje, Steve and Collier, Nigel and van der Linden, Sander and Roozenbeek, Jon},
title={Generative language models exhibit social identity biases},
journal={Nature Computational Science},
year={2025},
month={Jan},
day={01},
volume={5},
number={1},
pages={65-75},
issn={2662-8457},
doi={10.1038/s43588-024-00741-1},
url={https://doi.org/10.1038/s43588-024-00741-1}
}

@misc{wang2026agentshumansoutgroupbeliefdependent,
      title={When Agents See Humans as the Outgroup: Belief-Dependent Bias in LLM-Powered Agents}, 
      author={Zongwei Wang and Bincheng Gu and Hongyu Yu and Junliang Yu and Tao He and Jiayin Feng and Chenghua Lin and Min Gao},
      year={2026},
      eprint={2601.00240},
      archivePrefix={arXiv},
      primaryClass={cs.AI},
      url={https://arxiv.org/abs/2601.00240}, 
}

@misc{choi2026identityskewsdebateanonymization,
      title={When Identity Skews Debate: Anonymization for Bias-Reduced Multi-Agent Reasoning}, 
      author={Hyeong Kyu Choi and Xiaojin Zhu and Sharon Li},
      year={2026},
      eprint={2510.07517},
      archivePrefix={arXiv},
      primaryClass={cs.AI},
      url={https://arxiv.org/abs/2510.07517}, 
}

@Article{Zomer2026-ja,
author={Zomer, Nicola
and De Domenico, Manlio},
title={Unraveling the emergence of collective behavior in networks of cognitive agents},
journal={npj Artificial Intelligence},
year={2026},
month={Mar},
day={21},
volume={2},
number={1},
pages={36},
issn={3005-1460},
doi={10.1038/s44387-026-00091-5},
url={https://doi.org/10.1038/s44387-026-00091-5}
}

@article{article,
    author = {Tajfel, Henri and Billig, Michael and Bundy, Robert and Flament, Claude},
    year = {1971},
    month = {04},
    pages = {149 - 178},
    title = {Social Categorization and Inter-Group Behavior},
    volume = {1},
    journal = {European Journal of Social Psychology},
    doi = {10.1002/ejsp.2420010202}
}

@article{tajfel1970experiments,
 ISSN = {00368733, 19467087},
 URL = {http://www.jstor.org/stable/24927662},
 author = {Henri Tajfel},
 journal = {Scientific American},
 number = {5},
 pages = {96--103},
 publisher = {Scientific American, a division of Nature America, Inc.},
 title = {Experiments in Intergroup Discrimination},
 urldate = {2026-05-06},
 volume = {223},
 year = {1970}
}

@incollection{tajfel1979integrative,
  title     = {An Integrative Theory of Intergroup Conflict},
  author    = {Tajfel, Henri and Turner, John C.},
  booktitle = {The Social Psychology of Intergroup Relations},
  editor    = {Austin, William G. and Worchel, Stephen},
  publisher = {Brooks/Cole},
  address   = {Monterey, CA},
  pages     = {33--47},
  year      = {1979}
}

@article{brewer1979ingroup,
  title={In-group bias in the minimal intergroup situation: A cognitive-motivational analysis.},
  author={Marilynn B. Brewer},
  journal={Psychological Bulletin},
  year={1979},
  volume={86},
  pages={307-324},
  url={https://api.semanticscholar.org/CorpusID:146445745}
}

@inproceedings{cemri2025multiagentllmsystemsfail,
title={Why Do Multi-Agent {LLM} Systems Fail?},
author={Mert Cemri and Melissa Z Pan and Shuyi Yang and Lakshya A Agrawal and Bhavya Chopra and Rishabh Tiwari and Kurt Keutzer and Aditya Parameswaran and Dan Klein and Kannan Ramchandran and Matei Zaharia and Joseph E. Gonzalez and Ion Stoica},
booktitle={The Thirty-ninth Annual Conference on Neural Information Processing Systems Datasets and Benchmarks Track},
year={2026},
url={https://openreview.net/forum?id=fAjbYBmonr}
}

@article{meta2022cicero,
author = {Anton Bakhtin  and Noam Brown  and Emily Dinan  and Gabriele Farina  and Colin Flaherty  and Daniel Fried  and Andrew Goff  and Jonathan Gray  and Hengyuan Hu  and Athul Paul Jacob  and Mojtaba Komeili  and Karthik Konath  and Minae Kwon  and Adam Lerer  and Mike Lewis  and Alexander H. Miller  and Sasha Mitts  and Adithya Renduchintala  and Stephen Roller  and Dirk Rowe  and Weiyan Shi  and Joe Spisak  and Alexander Wei  and David Wu  and Hugh Zhang  and Markus Zijlstra },
title = {Human-level play in the game of Diplomacy by combining language models with strategic reasoning},
journal = {Science},
volume = {378},
number = {6624},
pages = {1067-1074},
year = {2022},
doi = {10.1126/science.ade9097},
URL = {https://www.science.org/doi/abs/10.1126/science.ade9097},
eprint = {https://www.science.org/doi/pdf/10.1126/science.ade9097},
}

@inproceedings{zhu2025multiagentbenchevaluatingcollaborationcompetition,
    title = "{M}ulti{A}gent{B}ench : Evaluating the Collaboration and Competition of {LLM} agents",
    author = "Zhu, Kunlun  and
      Du, Hongyi  and
      Hong, Zhaochen  and
      Yang, Xiaocheng  and
      Guo, Shuyi  and
      Wang, Zhe  and
      Wang, Zhenhailong  and
      Qian, Cheng  and
      Tang, Xiangru  and
      Ji, Heng  and
      You, Jiaxuan",
    editor = "Che, Wanxiang  and
      Nabende, Joyce  and
      Shutova, Ekaterina  and
      Pilehvar, Mohammad Taher",
    booktitle = "Proceedings of the 63rd Annual Meeting of the Association for Computational Linguistics (Volume 1: Long Papers)",
    month = jul,
    year = "2025",
    address = "Vienna, Austria",
    publisher = "Association for Computational Linguistics",
    url = "https://aclanthology.org/2025.acl-long.421/",
    doi = "10.18653/v1/2025.acl-long.421",
    pages = "8580--8622",
    ISBN = "979-8-89176-251-0"
}

@misc{wang2024battleagentbenchbenchmarkevaluatingcooperation,
      title={BattleAgentBench: A Benchmark for Evaluating Cooperation and Competition Capabilities of Language Models in Multi-Agent Systems}, 
      author={Wei Wang and Dan Zhang and Tao Feng and Boyan Wang and Jie Tang},
      year={2024},
      eprint={2408.15971},
      archivePrefix={arXiv},
      primaryClass={cs.CL},
      url={https://arxiv.org/abs/2408.15971}, 
}

@article{birds,
 ISSN = {03600572, 15452115},
 URL = {http://www.jstor.org/stable/2678628},
 author = {Miller McPherson and Lynn Smith-Lovin and James M. Cook},
 journal = {Annual Review of Sociology},
 pages = {415--444},
 publisher = {Annual Reviews},
 title = {Birds of a Feather: Homophily in Social Networks},
 urldate = {2026-05-04},
 volume = {27},
 year = {2001}
}

@article{nash1950bargaining,
 ISSN = {00129682, 14680262},
 URL = {http://www.jstor.org/stable/1907266},
 author = {John F. Nash},
 journal = {Econometrica},
 number = {2},
 pages = {155--162},
 publisher = {[Wiley, Econometric Society]},
 title = {The Bargaining Problem},
 urldate = {2026-09-25},
 volume = {18},
 year = {1950}
}

@article{rubinstein1982perfect,
 ISSN = {00129682, 14680262},
 URL = {http://www.jstor.org/stable/1912531},
 author = {Ariel Rubinstein},
 journal = {Econometrica},
 number = {1},
 pages = {97--109},
 publisher = {[Wiley, Econometric Society]},
 title = {Perfect Equilibrium in a Bargaining Model},
 urldate = {2026-09-25},
 volume = {50},
 year = {1982}
}

@article{crawford1982strategic,
author = {Joel Sobel AND Vincent P. Crawford},
title = {Strategic Information Transmission},
journal = {Econometrica},
volume = {50},
number = {6},
pages = {1431-1451},
url = {https://onlinelibrary.wiley.com/doi/abs/10.2307/1913390},
year = {1982}
}

@inproceedings{abdelnabi2024cooperation,
 author = {Abdelnabi, Sahar and Gomaa, Amr and Sivaprasad, Sarath and Sch\"{o}nherr, Lea and Fritz, Mario},
 booktitle = {Advances in Neural Information Processing Systems},
 doi = {10.52202/079017-2658},
 editor = {A. Globerson and L. Mackey and D. Belgrave and A. Fan and U. Paquet and J. Tomczak and C. Zhang},
 pages = {83548--83599},
 publisher = {Curran Associates, Inc.},
 title = {Cooperation, Competition, and Maliciousness: LLM-Stakeholders Interactive Negotiation},
 url = {https://proceedings.neurips.cc/paper_files/paper/2024/file/984dd3db213db2d1454a163b65b84d08-Paper-Datasets_and_Benchmarks_Track.pdf},
 volume = {37},
 year = {2024}
}

@inproceedings{park2025maporl,
    title = "{MAP}o{RL}: Multi-Agent Post-Co-Training for Collaborative Large Language Models with Reinforcement Learning",
    author = "Park, Chanwoo  and
      Han, Seungju  and
      Guo, Xingzhi  and
      Ozdaglar, Asuman E.  and
      Zhang, Kaiqing  and
      Kim, Joo-Kyung",
    editor = "Che, Wanxiang  and
      Nabende, Joyce  and
      Shutova, Ekaterina  and
      Pilehvar, Mohammad Taher",
    booktitle = "Proceedings of the 63rd Annual Meeting of the Association for Computational Linguistics (Volume 1: Long Papers)",
    month = jul,
    year = "2025",
    address = "Vienna, Austria",
    publisher = "Association for Computational Linguistics",
    url = "https://aclanthology.org/2025.acl-long.1459/",
    doi = "10.18653/v1/2025.acl-long.1459",
    pages = "30215--30248",
    ISBN = "979-8-89176-251-0"
}

@article{akerlof2000economics,
 ISSN = {00335533, 15314650},
 URL = {http://www.jstor.org/stable/2586894},
 author = {George A. Akerlof and Rachel E. Kranton},
 journal = {The Quarterly Journal of Economics},
 number = {3},
 pages = {715--753},
 publisher = {Oxford University Press},
 title = {Economics and Identity},
 urldate = {2026-05-06},
 volume = {115},
 year = {2000}
}

@ARTICLE{chen2009group,
title = {Group Identity and Social Preferences},
author = {Chen, Yan and Li, Sherry},
year = {2009},
journal = {American Economic Review},
volume = {99},
number = {1},
pages = {431-57},
url = {https://EconPapers.repec.org/RePEc:aea:aecrev:v:99:y:2009:i:1:p:431-57}
}

@misc{faulkner2026elections,
      title={Evaluating Cooperation in {LLM} Social Groups through Elected Leadership}, 
      author={Ryan Faulkner and Anushka Deshpande and David Guzman Piedrahita and Joel Z. Leibo and Zhijing Jin},
      year={2026},
      eprint={2604.11721},
      archivePrefix={arXiv},
      primaryClass={cs.CL},
      url={https://arxiv.org/abs/2604.11721}, 
}

@misc{kumar2026cooperative,
      title={Cooperative Profiles Predict Multi-Agent {LLM} Team Performance in {AI} for Science Workflows}, 
      author={Shivani Kumar and Adarsh Bharathwaj and David Jurgens},
      year={2026},
      eprint={2604.20658},
      archivePrefix={arXiv},
      primaryClass={cs.CL},
      url={https://arxiv.org/abs/2604.20658}, 
}

@misc{migliarini2026quantifyingselfpreservationbiaslarge,
      title={Quantifying Self-Preservation Bias in Large Language Models}, 
      author={Matteo Migliarini and Joaquin Pereira Pizzini and Luca Moresca and Valerio Santini and Indro Spinelli and Fabio Galasso},
      year={2026},
      eprint={2604.02174},
      archivePrefix={arXiv},
      primaryClass={cs.AI},
      url={https://arxiv.org/abs/2604.02174}, 
}

@misc{metr2026huggingfaceincident,
    title = {Brief independent investigation of agents' behavior, reasoning and collaboration in the OpenAI / Hugging Face hacking incident},
    author = {METR},
    howpublished = {\url{https://metr.org/blog/2026-08-26-openai-hugging-face-incident-investigation/}},
    year = {2026},
    month = {08},
}

@misc{anthropic2026multiagentpatterns,
  author       = {{Anthropic Frontier Red Team}},
  title        = {Patterns and problems in emerging multiagent systems},
  year         = {2026},
  month        = aug,
  howpublished = {\url{https://www.anthropic.com/research/multiagent-systems}},
  note         = {Corresponding author Carolyn Zou. 13 August 2026. Accessed 2026-09-07.},
}

@inproceedings{rein2024gpqa,
title={{GPQA}: A Graduate-Level Google-Proof Q\&A Benchmark},
author={David Rein and Betty Li Hou and Asa Cooper Stickland and Jackson Petty and Richard Yuanzhe Pang and Julien Dirani and Julian Michael and Samuel R. Bowman},
booktitle={First Conference on Language Modeling},
year={2024},
url={https://openreview.net/forum?id=Ti67584b98}
}

@inproceedings{wu2024autogen,
title={AutoGen: Enabling Next-Gen {LLM} Applications via Multi-Agent Conversations},
author={Qingyun Wu and Gagan Bansal and Jieyu Zhang and Yiran Wu and Beibin Li and Erkang Zhu and Li Jiang and Xiaoyun Zhang and Shaokun Zhang and Jiale Liu and Ahmed Hassan Awadallah and Ryen W White and Doug Burger and Chi Wang},
booktitle={First Conference on Language Modeling},
year={2024},
url={https://openreview.net/forum?id=BAakY1hNKS}
}

@inproceedings{wang2025mixtureofagents,
title={Mixture-of-Agents Enhances Large Language Model Capabilities},
author={Junlin Wang and Jue WANG and Ben Athiwaratkun and Ce Zhang and James Zou},
booktitle={The Thirteenth International Conference on Learning Representations},
year={2025},
url={https://openreview.net/forum?id=h0ZfDIrj7T}
}

@inproceedings{chen2025internet,
title={Internet of Agents: Weaving a Web of Heterogeneous Agents for Collaborative Intelligence},
author={Weize Chen and Ziming You and Ran Li and Yitong Guan and Chen Qian and Chenyang Zhao and Cheng Yang and Ruobing Xie and Zhiyuan Liu and Maosong Sun},
booktitle={The Thirteenth International Conference on Learning Representations},
year={2025},
url={https://openreview.net/forum?id=o1Et3MogPw}
}

\clearpage
\appendix
\section{Additional Results}
\label{sec:appendix}

This appendix collects supporting results that are referenced only briefly in
the main text.

Tables~\ref{tab:factionalism-le-full}--\ref{tab:gpqa-full} report EDR next to AMI for every experiment of Section~\ref{sec:results}, under family labels, under neutral color labels and on GPQA-Diamond. EDR is the homophily measure of Section~\ref{sec:measurement}; bold marks the decision rule of the same section.

\begin{table}[htbp]
\centering
\caption{\iconLE~Leader Election factionalism, EDR and AMI, for the experiments of Table~\ref{tab:factionalism}. For \emph{mislabeled} runs the same trajectories are scored against the visible-label partition $\ell^{\mathrm{vis}}$ and the true-family partition $\ell^{\mathrm{true}}$. Subscripts are standard errors; significant results in \textbf{bold}.}
\label{tab:factionalism-le-full}
\setlength{\tabcolsep}{2.2pt}
\small
\begin{tabular}{llcccccc}
\toprule
\multirow{2}{*}{\textbf{Composition}} & \multirow{2}{*}{\textbf{Condition}} & \multicolumn{2}{c}{$\mathbf{N=9}$ (3 families)} & \multicolumn{2}{c}{$\mathbf{N=12}$ (4 families)} & \multicolumn{2}{c}{$\mathbf{N=25}$ (5 families)} \\
\cmidrule(lr){3-4} \cmidrule(lr){5-6} \cmidrule(lr){7-8}
& & EDR & AMI & EDR & AMI & EDR & AMI \\
\midrule
\multirow{4}{*}{\textbf{Unbalanced}}
 & Unlabeled                         & $1.10_{\pm 0.06}$ & $0.05_{\pm 0.03}$ & $1.01_{\pm 0.06}$ & $0.03_{\pm 0.02}$ & $1.03_{\pm 0.04}$ & $0.02_{\pm 0.01}$ \\
 & Labeled                           & $\mathbf{1.87}_{\pm 0.17}$ & $\mathbf{0.43}_{\pm 0.05}$ & $\mathbf{1.65}_{\pm 0.09}$ & $\mathbf{0.30}_{\pm 0.04}$ & $\mathbf{1.70}_{\pm 0.08}$ & $\mathbf{0.24}_{\pm 0.02}$ \\
 & Mislabeled $\ell^{\mathrm{vis}}$  & $\mathbf{1.54}_{\pm 0.12}$ & $\mathbf{0.33}_{\pm 0.05}$ & $\mathbf{1.55}_{\pm 0.09}$ & $\mathbf{0.18}_{\pm 0.03}$ & $\mathbf{1.56}_{\pm 0.07}$ & $\mathbf{0.21}_{\pm 0.02}$ \\
 & Mislabeled $\ell^{\mathrm{true}}$ & $\mathbf{1.27}_{\pm 0.07}$ & $\mathbf{0.18}_{\pm 0.03}$ & $0.98_{\pm 0.07}$ & $0.04_{\pm 0.03}$ & $1.09_{\pm 0.05}$ & $0.08_{\pm 0.01}$ \\
\midrule
\multirow{4}{*}{\textbf{Balanced}}
 & Unlabeled                         & $1.12_{\pm 0.07}$ & $0.08_{\pm 0.03}$ & $1.16_{\pm 0.06}$ & $0.06_{\pm 0.02}$ & $1.01_{\pm 0.04}$ & $0.03_{\pm 0.01}$ \\
 & Labeled                           & $\mathbf{2.46}_{\pm 0.20}$ & $\mathbf{0.53}_{\pm 0.04}$ & $\mathbf{1.74}_{\pm 0.11}$ & $\mathbf{0.31}_{\pm 0.03}$ & $\mathbf{1.78}_{\pm 0.06}$ & $\mathbf{0.28}_{\pm 0.02}$ \\
 & Mislabeled $\ell^{\mathrm{vis}}$  & $\mathbf{1.50}_{\pm 0.11}$ & $\mathbf{0.35}_{\pm 0.04}$ & $\mathbf{1.45}_{\pm 0.08}$ & $\mathbf{0.23}_{\pm 0.03}$ & $\mathbf{1.75}_{\pm 0.07}$ & $\mathbf{0.24}_{\pm 0.02}$ \\
 & Mislabeled $\ell^{\mathrm{true}}$ & $1.10_{\pm 0.07}$ & $0.10_{\pm 0.03}$ & $1.10_{\pm 0.08}$ & $0.04_{\pm 0.02}$ & $1.13_{\pm 0.07}$ & $0.07_{\pm 0.02}$ \\
\bottomrule
\end{tabular}
\end{table}

\begin{table}[htbp]
\centering
\caption{\iconEX~Exclusion factionalism, EDR and AMI, for the experiments of Table~\ref{tab:factionalism}. Conventions as in Table~\ref{tab:factionalism-le-full}. Significant results in \textbf{bold}.}
\label{tab:factionalism-ex-full}
\setlength{\tabcolsep}{2.2pt}
\small
\begin{tabular}{llcccccc}
\toprule
\multirow{2}{*}{\textbf{Composition}} & \multirow{2}{*}{\textbf{Condition}} & \multicolumn{2}{c}{$\mathbf{N=9}$ (3 families)} & \multicolumn{2}{c}{$\mathbf{N=12}$ (4 families)} & \multicolumn{2}{c}{$\mathbf{N=25}$ (5 families)} \\
\cmidrule(lr){3-4} \cmidrule(lr){5-6} \cmidrule(lr){7-8}
& & EDR & AMI & EDR & AMI & EDR & AMI \\
\midrule
\multirow{4}{*}{\textbf{Unbalanced}}
 & Unlabeled                         & $1.27_{\pm 0.09}$ & $0.07_{\pm 0.04}$ & $1.02_{\pm 0.07}$ & $0.08_{\pm 0.03}$ & $0.94_{\pm 0.05}$ & $0.02_{\pm 0.01}$ \\
 & Labeled                           & $\mathbf{3.44}_{\pm 0.23}$ & $\mathbf{0.65}_{\pm 0.03}$ & $\mathbf{3.06}_{\pm 0.21}$ & $\mathbf{0.59}_{\pm 0.03}$ & $\mathbf{3.78}_{\pm 0.24}$ & $\mathbf{0.55}_{\pm 0.02}$ \\
 & Mislabeled $\ell^{\mathrm{vis}}$  & $\mathbf{3.32}_{\pm 0.23}$ & $\mathbf{0.67}_{\pm 0.03}$ & $\mathbf{2.72}_{\pm 0.14}$ & $\mathbf{0.58}_{\pm 0.03}$ & $\mathbf{3.98}_{\pm 0.21}$ & $\mathbf{0.55}_{\pm 0.03}$ \\
 & Mislabeled $\ell^{\mathrm{true}}$ & $\mathbf{1.54}_{\pm 0.12}$ & $\mathbf{0.29}_{\pm 0.03}$ & $1.12_{\pm 0.07}$ & $0.11_{\pm 0.03}$ & $1.07_{\pm 0.05}$ & $0.08_{\pm 0.01}$ \\
\midrule
\multirow{4}{*}{\textbf{Balanced}}
 & Unlabeled                         & $0.94_{\pm 0.08}$ & $0.04_{\pm 0.04}$ & $1.04_{\pm 0.06}$ & $0.05_{\pm 0.03}$ & $1.06_{\pm 0.03}$ & $0.04_{\pm 0.01}$ \\
 & Labeled                           & $\mathbf{4.72}_{\pm 0.50}$ & $\mathbf{0.76}_{\pm 0.03}$ & $\mathbf{2.98}_{\pm 0.17}$ & $\mathbf{0.62}_{\pm 0.03}$ & $\mathbf{4.23}_{\pm 0.19}$ & $\mathbf{0.61}_{\pm 0.02}$ \\
 & Mislabeled $\ell^{\mathrm{vis}}$  & $\mathbf{2.95}_{\pm 0.18}$ & $\mathbf{0.69}_{\pm 0.04}$ & $\mathbf{2.49}_{\pm 0.14}$ & $\mathbf{0.58}_{\pm 0.03}$ & $\mathbf{4.32}_{\pm 0.18}$ & $\mathbf{0.59}_{\pm 0.02}$ \\
 & Mislabeled $\ell^{\mathrm{true}}$ & $1.15_{\pm 0.09}$ & $0.13_{\pm 0.03}$ & $1.09_{\pm 0.06}$ & $0.03_{\pm 0.02}$ & $1.10_{\pm 0.06}$ & $0.05_{\pm 0.01}$ \\
\bottomrule
\end{tabular}
\end{table}

\begin{table}[htbp]
\centering
\caption{\iconLE~Leader Election factionalism with neutral color labels, EDR and AMI, for the experiments of Table~\ref{tab:colors}. Unlabeled baselines are those of Table~\ref{tab:factionalism-le-full}. Significant results in \textbf{bold}.}
\label{tab:colors-le-full}
\setlength{\tabcolsep}{2.2pt}
\small
\begin{tabular}{llcccccc}
\toprule
\multirow{2}{*}{\textbf{Composition}} & \multirow{2}{*}{\textbf{Condition}} & \multicolumn{2}{c}{$\mathbf{N=9}$ (3 families)} & \multicolumn{2}{c}{$\mathbf{N=12}$ (4 families)} & \multicolumn{2}{c}{$\mathbf{N=25}$ (5 families)} \\
\cmidrule(lr){3-4} \cmidrule(lr){5-6} \cmidrule(lr){7-8}
& & EDR & AMI & EDR & AMI & EDR & AMI \\
\midrule
\multirow{3}{*}{\textbf{Unbalanced}}
 & Labeled                           & $\mathbf{2.07}_{\pm 0.20}$ & $\mathbf{0.37}_{\pm 0.05}$ & $\mathbf{1.48}_{\pm 0.09}$ & $\mathbf{0.21}_{\pm 0.03}$ & $\mathbf{1.65}_{\pm 0.07}$ & $\mathbf{0.24}_{\pm 0.02}$ \\
 & Mislabeled $\ell^{\mathrm{vis}}$  & $\mathbf{1.62}_{\pm 0.12}$ & $\mathbf{0.31}_{\pm 0.04}$ & $\mathbf{1.39}_{\pm 0.10}$ & $\mathbf{0.19}_{\pm 0.03}$ & $\mathbf{1.44}_{\pm 0.07}$ & $\mathbf{0.18}_{\pm 0.02}$ \\
 & Mislabeled $\ell^{\mathrm{true}}$ & $0.93_{\pm 0.08}$ & $0.01_{\pm 0.03}$ & $1.08_{\pm 0.06}$ & $0.02_{\pm 0.02}$ & $1.10_{\pm 0.05}$ & $0.07_{\pm 0.01}$ \\
\midrule
\multirow{3}{*}{\textbf{Balanced}}
 & Labeled                           & $\mathbf{1.81}_{\pm 0.10}$ & $\mathbf{0.43}_{\pm 0.04}$ & $\mathbf{1.56}_{\pm 0.12}$ & $\mathbf{0.25}_{\pm 0.03}$ & $\mathbf{2.01}_{\pm 0.09}$ & $\mathbf{0.26}_{\pm 0.02}$ \\
 & Mislabeled $\ell^{\mathrm{vis}}$  & $\mathbf{1.37}_{\pm 0.08}$ & $\mathbf{0.28}_{\pm 0.04}$ & $\mathbf{1.51}_{\pm 0.09}$ & $\mathbf{0.21}_{\pm 0.03}$ & $\mathbf{1.68}_{\pm 0.08}$ & $\mathbf{0.21}_{\pm 0.02}$ \\
 & Mislabeled $\ell^{\mathrm{true}}$ & $1.02_{\pm 0.07}$ & $-0.03_{\pm 0.03}$ & $1.17_{\pm 0.09}$ & $0.06_{\pm 0.02}$ & $1.08_{\pm 0.05}$ & $0.06_{\pm 0.01}$ \\
\bottomrule
\end{tabular}
\end{table}

\begin{table}[htbp]
\centering
\caption{\iconEX~Exclusion factionalism with neutral color labels, EDR and AMI, for the experiments of Table~\ref{tab:colors}. Unlabeled baselines are those of Table~\ref{tab:factionalism-ex-full}. Significant results in \textbf{bold}.}
\label{tab:colors-ex-full}
\setlength{\tabcolsep}{2.2pt}
\small
\begin{tabular}{llcccccc}
\toprule
\multirow{2}{*}{\textbf{Composition}} & \multirow{2}{*}{\textbf{Condition}} & \multicolumn{2}{c}{$\mathbf{N=9}$ (3 families)} & \multicolumn{2}{c}{$\mathbf{N=12}$ (4 families)} & \multicolumn{2}{c}{$\mathbf{N=25}$ (5 families)} \\
\cmidrule(lr){3-4} \cmidrule(lr){5-6} \cmidrule(lr){7-8}
& & EDR & AMI & EDR & AMI & EDR & AMI \\
\midrule
\multirow{3}{*}{\textbf{Unbalanced}}
 & Labeled                           & $\mathbf{3.53}_{\pm 0.31}$ & $\mathbf{0.66}_{\pm 0.03}$ & $\mathbf{3.42}_{\pm 0.18}$ & $\mathbf{0.69}_{\pm 0.03}$ & $\mathbf{6.37}_{\pm 0.34}$ & $\mathbf{0.80}_{\pm 0.02}$ \\
 & Mislabeled $\ell^{\mathrm{vis}}$  & $\mathbf{3.97}_{\pm 0.25}$ & $\mathbf{0.76}_{\pm 0.02}$ & $\mathbf{3.85}_{\pm 0.20}$ & $\mathbf{0.72}_{\pm 0.03}$ & $\mathbf{6.98}_{\pm 0.48}$ & $\mathbf{0.79}_{\pm 0.02}$ \\
 & Mislabeled $\ell^{\mathrm{true}}$ & $0.72_{\pm 0.05}$ & $-0.12_{\pm 0.02}$ & $0.99_{\pm 0.07}$ & $-0.02_{\pm 0.02}$ & $1.02_{\pm 0.05}$ & $-0.01_{\pm 0.01}$ \\
\midrule
\multirow{3}{*}{\textbf{Balanced}}
 & Labeled                           & $\mathbf{5.15}_{\pm 0.46}$ & $\mathbf{0.82}_{\pm 0.03}$ & $\mathbf{3.58}_{\pm 0.20}$ & $\mathbf{0.67}_{\pm 0.03}$ & $\mathbf{8.09}_{\pm 0.56}$ & $\mathbf{0.81}_{\pm 0.02}$ \\
 & Mislabeled $\ell^{\mathrm{vis}}$  & $\mathbf{5.66}_{\pm 0.48}$ & $\mathbf{0.95}_{\pm 0.02}$ & $\mathbf{3.38}_{\pm 0.18}$ & $\mathbf{0.75}_{\pm 0.03}$ & $\mathbf{8.05}_{\pm 0.51}$ & $\mathbf{0.84}_{\pm 0.02}$ \\
 & Mislabeled $\ell^{\mathrm{true}}$ & $0.88_{\pm 0.07}$ & $-0.05_{\pm 0.02}$ & $0.95_{\pm 0.05}$ & $-0.01_{\pm 0.02}$ & $1.17_{\pm 0.05}$ & $0.01_{\pm 0.01}$ \\
\bottomrule
\end{tabular}
\end{table}

\begin{table}[htbp]
\centering
\caption{\iconGQ~GPQA-Diamond factionalism, EDR and AMI, for the experiments of Table~\ref{tab:gpqa}. Conventions as in Table~\ref{tab:factionalism-le-full}. Significant results in \textbf{bold}.}
\label{tab:gpqa-full}
\setlength{\tabcolsep}{2.2pt}
\small
\begin{tabular}{llcccc}
\toprule
\multirow{2}{*}{\textbf{Composition}} & \multirow{2}{*}{\textbf{Condition}} & \multicolumn{2}{c}{$\mathbf{N=9}$ (3 families)} & \multicolumn{2}{c}{$\mathbf{N=12}$ (4 families)} \\
\cmidrule(lr){3-4} \cmidrule(lr){5-6}
& & EDR & AMI & EDR & AMI \\
\midrule
\multirow{4}{*}{\textbf{Unbalanced}}
 & Unlabeled                          & $0.89_{\pm 0.04}$ & $0.05_{\pm 0.02}$ & $0.77_{\pm 0.03}$ & $0.00_{\pm 0.01}$ \\
 & Labeled                            & $\mathbf{1.66}_{\pm 0.08}$ & $\mathbf{0.28}_{\pm 0.02}$ & $\mathbf{1.98}_{\pm 0.08}$ & $\mathbf{0.32}_{\pm 0.02}$ \\
 & Mislabeled $\ell^{\mathrm{vis}}$   & $\mathbf{1.69}_{\pm 0.07}$ & $\mathbf{0.27}_{\pm 0.02}$ & $\mathbf{1.43}_{\pm 0.05}$ & $\mathbf{0.17}_{\pm 0.02}$ \\
 & Mislabeled $\ell^{\mathrm{true}}$  & $\mathbf{1.35}_{\pm 0.05}$ & $\mathbf{0.17}_{\pm 0.02}$ & $1.12_{\pm 0.07}$ & $0.07_{\pm 0.01}$ \\
\midrule
\multirow{4}{*}{\textbf{Balanced}}
 & Unlabeled                          & $0.78_{\pm 0.03}$ & $0.02_{\pm 0.01}$ & $0.80_{\pm 0.03}$ & $0.02_{\pm 0.01}$ \\
 & Labeled                            & $\mathbf{2.08}_{\pm 0.09}$ & $\mathbf{0.40}_{\pm 0.02}$ & $\mathbf{2.13}_{\pm 0.08}$ & $\mathbf{0.35}_{\pm 0.02}$ \\
 & Mislabeled $\ell^{\mathrm{vis}}$   & $\mathbf{1.73}_{\pm 0.06}$ & $\mathbf{0.30}_{\pm 0.02}$ & $\mathbf{2.06}_{\pm 0.08}$ & $\mathbf{0.31}_{\pm 0.01}$ \\
 & Mislabeled $\ell^{\mathrm{true}}$  & $1.15_{\pm 0.09}$ & $0.14_{\pm 0.02}$ & $1.10_{\pm 0.06}$ & $0.13_{\pm 0.01}$ \\
\bottomrule
\end{tabular}
\end{table}

Tables~\ref{tab:mechanism-e4-le} and~\ref{tab:mechanism-e4-ex} give the
family-resolved adoption-gap values discussed in Section~\ref{sec:results-mechanism}. Unlike the experiment-level EDR and AMI
statistics, these measurements operate directly at the token-decision surface, so
they help localize whether the preference for within-group continuations is
driven by the announced label or by the true architecture. Each entry gives the
within-family minus between-family gap for the adoption continuation
\texttt{VOTE: $v_j^{(r)}$}. In \emph{Unlabeled}, \emph{Labeled}, and
\emph{Mislabeled (true family)}, families are the true architectural ones; in
\emph{Mislabeled (visible label)}, they are the labels shown in the prompt.
Under \emph{Labeled}, the two partitions coincide, so only one block is shown;
under \emph{Unlabeled}, no visible-label block exists. Subscripts are
bootstrap standard errors over trajectories (10\,000 resamples), and
\textbf{bold} entries have 95\% confidence intervals that exclude zero.

\begin{table}[htbp]
\centering
\setlength{\tabcolsep}{3pt}
\renewcommand{\arraystretch}{1.15}
\caption{\iconLE~Leader Election adoption gap, by the evaluator's underlying model family.}
\label{tab:mechanism-e4-le}
\small
\begin{tabular}{llcccc}
\toprule
\multirow{1}{*}{\textbf{Condition}} & \multirow{1}{*}{\textbf{Composition}} & \textbf{Gemma} & \textbf{GPT-OSS} & \textbf{Nemotron} & \textbf{Qwen} \\
\midrule
\multirow{2}{*}{Unlabeled} & Balanced & $\mathbf{-0.390}_{\pm 0.082}$ & $\mathbf{+0.100}_{\pm 0.019}$ & $\mathbf{+0.036}_{\pm 0.018}$ & $+0.018_{\pm 0.028}$ \\
 & Unbalanced & $\mathbf{-0.396}_{\pm 0.071}$ & $\mathbf{+0.084}_{\pm 0.021}$ & $\mathbf{+0.046}_{\pm 0.015}$ & $-0.022_{\pm 0.024}$ \\
\midrule
\multirow{2}{*}{Labeled} & Balanced & $\mathbf{+0.844}_{\pm 0.108}$ & $\mathbf{+0.230}_{\pm 0.017}$ & $\mathbf{+0.119}_{\pm 0.023}$ & $\mathbf{+0.355}_{\pm 0.036}$ \\
 & Unbalanced & $\mathbf{+0.802}_{\pm 0.084}$ & $\mathbf{+0.247}_{\pm 0.032}$ & $\mathbf{+0.100}_{\pm 0.022}$ & $\mathbf{+0.304}_{\pm 0.030}$ \\
\midrule
\multirow{2}{*}{Mislabeled (true family)} & Balanced & $\mathbf{-0.324}_{\pm 0.077}$ & $\mathbf{+0.064}_{\pm 0.025}$ & $+0.013_{\pm 0.021}$ & $\mathbf{+0.082}_{\pm 0.026}$ \\
 & Unbalanced & $-0.102_{\pm 0.062}$ & $+0.033_{\pm 0.019}$ & $+0.013_{\pm 0.025}$ & $+0.054_{\pm 0.030}$ \\
\midrule
\multirow{2}{*}{Mislabeled (visible label)} & Balanced & $\mathbf{+0.666}_{\pm 0.055}$ & $\mathbf{+0.158}_{\pm 0.020}$ & $\mathbf{+0.120}_{\pm 0.015}$ & $\mathbf{+0.223}_{\pm 0.023}$ \\
 & Unbalanced & $\mathbf{+0.629}_{\pm 0.092}$ & $\mathbf{+0.195}_{\pm 0.019}$ & $\mathbf{+0.105}_{\pm 0.023}$ & $\mathbf{+0.251}_{\pm 0.023}$ \\
\bottomrule
\end{tabular}
\end{table}

\begin{table}[htbp]
\centering
\setlength{\tabcolsep}{3pt}
\renewcommand{\arraystretch}{1.15}
\caption{\iconEX~Exclusion adoption gap, by the evaluator's underlying model family.}
\label{tab:mechanism-e4-ex}
\small
\begin{tabular}{llcccc}
\toprule
\multirow{1}{*}{\textbf{Condition}} & \multirow{1}{*}{\textbf{Composition}} & \textbf{Gemma} & \textbf{GPT-OSS} & \textbf{Nemotron} & \textbf{Qwen} \\
\midrule
\multirow{2}{*}{Unlabeled} & Balanced & $\mathbf{+0.151}_{\pm 0.048}$ & $+0.025_{\pm 0.022}$ & $-0.008_{\pm 0.023}$ & $-0.017_{\pm 0.026}$ \\
 & Unbalanced & $\mathbf{+0.229}_{\pm 0.042}$ & $-0.004_{\pm 0.015}$ & $+0.002_{\pm 0.022}$ & $+0.011_{\pm 0.024}$ \\
\midrule
\multirow{2}{*}{Labeled} & Balanced & $\mathbf{+1.126}_{\pm 0.061}$ & $\mathbf{-0.074}_{\pm 0.033}$ & $\mathbf{+0.059}_{\pm 0.024}$ & $\mathbf{+0.180}_{\pm 0.028}$ \\
 & Unbalanced & $\mathbf{+1.164}_{\pm 0.056}$ & $+0.004_{\pm 0.027}$ & $\mathbf{+0.122}_{\pm 0.037}$ & $\mathbf{+0.207}_{\pm 0.023}$ \\
\midrule
\multirow{2}{*}{Mislabeled (true family)} & Balanced & $+0.112_{\pm 0.057}$ & $+0.022_{\pm 0.016}$ & $-0.006_{\pm 0.021}$ & $-0.003_{\pm 0.024}$ \\
 & Unbalanced & $\mathbf{+0.143}_{\pm 0.069}$ & $+0.003_{\pm 0.022}$ & $+0.003_{\pm 0.019}$ & $\mathbf{+0.147}_{\pm 0.029}$ \\
\midrule
\multirow{2}{*}{Mislabeled (visible label)} & Balanced & $\mathbf{+0.605}_{\pm 0.057}$ & $\mathbf{+0.131}_{\pm 0.021}$ & $\mathbf{+0.086}_{\pm 0.018}$ & $\mathbf{+0.206}_{\pm 0.021}$ \\
 & Unbalanced & $\mathbf{+0.610}_{\pm 0.062}$ & $\mathbf{+0.113}_{\pm 0.024}$ & $\mathbf{+0.076}_{\pm 0.021}$ & $\mathbf{+0.241}_{\pm 0.024}$ \\
\bottomrule
\end{tabular}
\end{table}

Tables~\ref{tab:nogemma-le} and~\ref{tab:nogemma-ex} report graph-level factionalism on Gemma-free rosters, testing whether the effect depends on a single family. At $N=9$ the roster is GPT-OSS, Nemotron and Qwen with Gemma removed; at $N=12$ Gemma is replaced by Seed-OSS (\ckpt{Seed-OSS-36B-Instruct}). Conventions follow Table~\ref{tab:factionalism}: for \emph{mislabeled} runs the same trajectories are scored against the visible-label partition $\ell^{\mathrm{vis}}$ and the true-family partition $\ell^{\mathrm{true}}$, and \textbf{bold} marks experiments passing all four corrected tests. Across the eight Gemma-free experiments the pattern of the main design is reproduced: unlabel remains a null in every experiment, labeling produces factionalism in all four experiments of each game, and under mislabeling the structure again follows the announced label rather than the true architecture. The effect is therefore not carried by a single family.

\begin{table}[htbp]
\centering
\caption{\iconLE~Leader Election factionalism on the agreement graph with Gemma removed from the roster. Significant results in \textbf{bold}.}
\label{tab:nogemma-le}
\setlength{\tabcolsep}{2.2pt}
\small
\begin{tabular}{llcccc}
\toprule
\multirow{2}{*}{\textbf{Composition}} & \multirow{2}{*}{\textbf{Condition}} & \multicolumn{2}{c}{$\mathbf{N=9}$ (3 families)} & \multicolumn{2}{c}{$\mathbf{N=12}$ (4 families)} \\
\cmidrule(lr){3-4} \cmidrule(lr){5-6}
& & EDR & AMI & EDR & AMI \\
\midrule
\multirow{4}{*}{\textbf{Unbalanced}}
 & Unlabeled                         & $1.04_{\pm 0.06}$ & $-0.01_{\pm 0.03}$ & $1.09_{\pm 0.05}$ & $+0.03_{\pm 0.03}$ \\
 & Labeled                           & $\mathbf{1.46}_{\pm 0.14}$ & $\mathbf{+0.24}_{\pm 0.05}$ & $\mathbf{1.39}_{\pm 0.06}$ & $\mathbf{+0.23}_{\pm 0.03}$ \\
 & Mislabeled $\ell^{\mathrm{vis}}$  & $\mathbf{1.13}_{\pm 0.06}$ & $\mathbf{+0.10}_{\pm 0.04}$ & $\mathbf{1.15}_{\pm 0.06}$ & $\mathbf{+0.10}_{\pm 0.03}$ \\
 & Mislabeled $\ell^{\mathrm{true}}$ & $1.02_{\pm 0.08}$ & $+0.05_{\pm 0.04}$ & $1.02_{\pm 0.05}$ & $+0.06_{\pm 0.02}$ \\
\midrule
\multirow{4}{*}{\textbf{Balanced}}
 & Unlabeled                         & $1.00_{\pm 0.06}$ & $+0.03_{\pm 0.04}$ & $1.02_{\pm 0.04}$ & $-0.03_{\pm 0.02}$ \\
 & Labeled                           & $\mathbf{1.40}_{\pm 0.07}$ & $\mathbf{+0.27}_{\pm 0.04}$ & $\mathbf{1.57}_{\pm 0.10}$ & $\mathbf{+0.22}_{\pm 0.03}$ \\
 & Mislabeled $\ell^{\mathrm{vis}}$  & $\mathbf{1.30}_{\pm 0.09}$ & $\mathbf{+0.23}_{\pm 0.04}$ & $\mathbf{1.21}_{\pm 0.07}$ & $\mathbf{+0.10}_{\pm 0.03}$ \\
 & Mislabeled $\ell^{\mathrm{true}}$ & $1.14_{\pm 0.05}$ & $+0.06_{\pm 0.03}$ & $0.98_{\pm 0.05}$ & $+0.03_{\pm 0.03}$ \\
\bottomrule
\end{tabular}
\end{table}

\begin{table}[htbp]
\centering
\caption{\iconEX~Exclusion factionalism on the agreement graph with Gemma removed from the roster. Rosters and conventions as in Table~\ref{tab:nogemma-le}. Significant results in \textbf{bold}.}
\label{tab:nogemma-ex}
\setlength{\tabcolsep}{2.2pt}
\small
\begin{tabular}{llcccc}
\toprule
\multirow{2}{*}{\textbf{Composition}} & \multirow{2}{*}{\textbf{Condition}} & \multicolumn{2}{c}{$\mathbf{N=9}$ (3 families)} & \multicolumn{2}{c}{$\mathbf{N=12}$ (4 families)} \\
\cmidrule(lr){3-4} \cmidrule(lr){5-6}
& & EDR & AMI & EDR & AMI \\
\midrule
\multirow{4}{*}{\textbf{Unbalanced}}
 & Unlabeled                         & $0.89_{\pm 0.05}$ & $+0.02_{\pm 0.03}$ & $1.04_{\pm 0.05}$ & $-0.01_{\pm 0.03}$ \\
 & Labeled                           & $\mathbf{2.26}_{\pm 0.18}$ & $\mathbf{+0.54}_{\pm 0.04}$ & $\mathbf{2.09}_{\pm 0.12}$ & $\mathbf{+0.42}_{\pm 0.04}$ \\
 & Mislabeled $\ell^{\mathrm{vis}}$  & $\mathbf{1.57}_{\pm 0.13}$ & $\mathbf{+0.32}_{\pm 0.05}$ & $\mathbf{1.92}_{\pm 0.11}$ & $\mathbf{+0.39}_{\pm 0.04}$ \\
 & Mislabeled $\ell^{\mathrm{true}}$ & $\mathbf{1.23}_{\pm 0.07}$ & $\mathbf{+0.14}_{\pm 0.04}$ & $1.03_{\pm 0.06}$ & $+0.04_{\pm 0.02}$ \\
\midrule
\multirow{4}{*}{\textbf{Balanced}}
 & Unlabeled                         & $0.95_{\pm 0.07}$ & $+0.08_{\pm 0.04}$ & $0.99_{\pm 0.05}$ & $+0.06_{\pm 0.03}$ \\
 & Labeled                           & $\mathbf{1.98}_{\pm 0.17}$ & $\mathbf{+0.53}_{\pm 0.05}$ & $\mathbf{2.06}_{\pm 0.11}$ & $\mathbf{+0.40}_{\pm 0.04}$ \\
 & Mislabeled $\ell^{\mathrm{vis}}$  & $\mathbf{1.83}_{\pm 0.12}$ & $\mathbf{+0.46}_{\pm 0.04}$ & $\mathbf{1.88}_{\pm 0.11}$ & $\mathbf{+0.39}_{\pm 0.04}$ \\
 & Mislabeled $\ell^{\mathrm{true}}$ & $1.05_{\pm 0.07}$ & $+0.07_{\pm 0.03}$ & $0.96_{\pm 0.07}$ & $+0.02_{\pm 0.02}$ \\
\bottomrule
\end{tabular}
\end{table}

Tables~\ref{tab:mechanism-e4-le-seedoss} and~\ref{tab:mechanism-e4-ex-seedoss} repeat the
same per-family decomposition on the Gemma-free roster at $N=12$, where Gemma is replaced
by Seed-OSS (\ckpt{Seed-OSS-36B-Instruct}), testing whether the
adoption effect depends on a single family. Conventions are identical to
Tables~\ref{tab:mechanism-e4-le} and~\ref{tab:mechanism-e4-ex}.

\begin{table}[htbp]
\centering
\setlength{\tabcolsep}{3pt}
\renewcommand{\arraystretch}{1.15}
\caption{\iconLE~Leader Election adoption gap by the evaluator's underlying model family, with Gemma replaced by Seed-OSS.}
\label{tab:mechanism-e4-le-seedoss}
\small
\begin{tabular}{llcccc}
\toprule
\multirow{1}{*}{\textbf{Condition}} & \multirow{1}{*}{\textbf{Composition}} & \textbf{GPT-OSS} & \textbf{Nemotron} & \textbf{Qwen} & \textbf{Seed-OSS} \\
\midrule
\multirow{2}{*}{Unlabeled} & Balanced & $-0.014_{\pm 0.019}$ & $\mathbf{-0.065}_{\pm 0.014}$ & $-0.001_{\pm 0.030}$ & $\mathbf{+0.068}_{\pm 0.022}$ \\
 & Unbalanced & $+0.008_{\pm 0.016}$ & $+0.007_{\pm 0.017}$ & $-0.069_{\pm 0.037}$ & $-0.016_{\pm 0.020}$ \\
\midrule
\multirow{2}{*}{Labeled} & Balanced & $\mathbf{+0.137}_{\pm 0.022}$ & $-0.013_{\pm 0.020}$ & $\mathbf{+0.265}_{\pm 0.036}$ & $\mathbf{+0.221}_{\pm 0.037}$ \\
 & Unbalanced & $\mathbf{+0.116}_{\pm 0.019}$ & $+0.032_{\pm 0.027}$ & $\mathbf{+0.205}_{\pm 0.039}$ & $\mathbf{+0.159}_{\pm 0.040}$ \\
\midrule
\multirow{2}{*}{Mislabeled (true family)} & Balanced & $+0.032_{\pm 0.018}$ & $\mathbf{-0.062}_{\pm 0.025}$ & $-0.002_{\pm 0.034}$ & $\mathbf{+0.101}_{\pm 0.033}$ \\
 & Unbalanced & $-0.021_{\pm 0.016}$ & $-0.017_{\pm 0.035}$ & $+0.018_{\pm 0.026}$ & $\mathbf{+0.104}_{\pm 0.038}$ \\
\midrule
\multirow{2}{*}{Mislabeled (visible label)} & Balanced & $\mathbf{+0.072}_{\pm 0.013}$ & $\mathbf{+0.039}_{\pm 0.015}$ & $\mathbf{+0.108}_{\pm 0.033}$ & $\mathbf{+0.117}_{\pm 0.027}$ \\
 & Unbalanced & $\mathbf{+0.055}_{\pm 0.021}$ & $\mathbf{+0.045}_{\pm 0.021}$ & $\mathbf{+0.118}_{\pm 0.018}$ & $+0.045_{\pm 0.033}$ \\
\bottomrule
\end{tabular}
\end{table}

\begin{table}[htbp]
\centering
\setlength{\tabcolsep}{3pt}
\renewcommand{\arraystretch}{1.15}
\caption{\iconEX~Exclusion adoption gap by the evaluator's underlying model family, with Gemma replaced by Seed-OSS.}
\label{tab:mechanism-e4-ex-seedoss}
\small
\begin{tabular}{llcccc}
\toprule
\multirow{1}{*}{\textbf{Condition}} & \multirow{1}{*}{\textbf{Composition}} & \textbf{GPT-OSS} & \textbf{Nemotron} & \textbf{Qwen} & \textbf{Seed-OSS} \\
\midrule
\multirow{2}{*}{Unlabeled} & Balanced & $\mathbf{+0.025}_{\pm 0.012}$ & $-0.011_{\pm 0.022}$ & $\mathbf{+0.042}_{\pm 0.018}$ & $\mathbf{+0.059}_{\pm 0.028}$ \\
 & Unbalanced & $+0.011_{\pm 0.024}$ & $+0.000_{\pm 0.014}$ & $-0.023_{\pm 0.018}$ & $+0.062_{\pm 0.043}$ \\
\midrule
\multirow{2}{*}{Labeled} & Balanced & $\mathbf{+0.106}_{\pm 0.038}$ & $+0.019_{\pm 0.027}$ & $\mathbf{+0.223}_{\pm 0.021}$ & $\mathbf{+0.162}_{\pm 0.034}$ \\
 & Unbalanced & $\mathbf{+0.264}_{\pm 0.093}$ & $\mathbf{+0.088}_{\pm 0.026}$ & $\mathbf{+0.253}_{\pm 0.025}$ & $\mathbf{+0.252}_{\pm 0.036}$ \\
\midrule
\multirow{2}{*}{Mislabeled (true family)} & Balanced & $\mathbf{-0.048}_{\pm 0.028}$ & $-0.002_{\pm 0.018}$ & $+0.020_{\pm 0.023}$ & $+0.046_{\pm 0.024}$ \\
 & Unbalanced & $\mathbf{+0.053}_{\pm 0.029}$ & $\mathbf{+0.050}_{\pm 0.021}$ & $+0.024_{\pm 0.021}$ & $+0.045_{\pm 0.031}$ \\
\midrule
\multirow{2}{*}{Mislabeled (visible label)} & Balanced & $\mathbf{+0.100}_{\pm 0.021}$ & $+0.036_{\pm 0.024}$ & $\mathbf{+0.187}_{\pm 0.021}$ & $\mathbf{+0.141}_{\pm 0.020}$ \\
 & Unbalanced & $\mathbf{+0.131}_{\pm 0.020}$ & $\mathbf{+0.073}_{\pm 0.018}$ & $\mathbf{+0.168}_{\pm 0.023}$ & $\mathbf{+0.157}_{\pm 0.019}$ \\
\bottomrule
\end{tabular}
\end{table}

\FloatBarrier

\newpage
\section{Measurement framework: full specification}
\label{app:inference}
This appendix gives the full notation, per-run hypothesis tests, and
batch-level inference machinery summarized in
Section~\ref{sec:measurement}.

\paragraph{Notation.} Consider a single game run with $n$ agents
$\mathcal{A} = \{a_1, \dots, a_n\}$ divided under the tested partition into $K$ groups
$\{G_1, \dots, G_K\}$ with sizes $n_1, \dots, n_K$. Each agent $a_i$
belongs to one of the groups $G_1, \dots, G_K$. The
within- and between-group pair sets defined in
Section~\ref{sec:measurement} have cardinalities
\begin{equation}
|\mathcal{W}| = \sum_{k=1}^{K} n_k(n_k - 1), \qquad
|\mathcal{B}| = n(n-1) - |\mathcal{W}|.
\end{equation}

\paragraph{Per-run homophily statistics.} The within-group and
between-group mean edge weights summarized in
Section~\ref{sec:measurement} are
\begin{equation}
d_W = \frac{\sum_{(i,j) \in \mathcal{W}} w(a_i \to a_j)}{|\mathcal{W}|},
\qquad
d_B = \frac{\sum_{(i,j) \in \mathcal{B}} w(a_i \to a_j)}{|\mathcal{B}|}.
\end{equation}

\paragraph{Per-run permutation test.} For each run we generate $B$
permutations $\pi_b$ of the group labels over $\mathcal{A}$. Because
each permutation is a reshuffle of the existing multiset of labels,
group-size marginals are preserved automatically. We recompute
$\mathrm{EDR}_b$ under each $\pi_b$ and report the one-sided $p$-value
\begin{equation}
p = \frac{1 + \sum_b \mathds{1}[\mathrm{EDR}_b \geq \mathrm{EDR}_{\mathrm{obs}}]}{B + 1},
\end{equation}
with $B = 1000$. The observed arrangement is counted among the $B+1$ equally likely ones, so $p$ is never zero and is bounded below by $1/(B+1)$; Fisher's combination below is therefore always defined. The test is distribution-free and requires no parametric assumption on
edge weights.

\paragraph{Absent edges and degenerate ratios.} An ordered pair $(a_i,a_j)$ for which $a_i$
never selected $a_j$ as a recipient offers no opportunity for agreement, so
Equation~\ref{eq:edge_construction} leaves that edge absent from the graph. The pair still belongs
to $\mathcal{W}$ or $\mathcal{B}$, whose sizes count every ordered pair of the partition, so in the
two densities it contributes a weight of zero. A pair that was contacted but never produced
agreement is a present edge with weight zero. 
A contact whose recipient is no longer active in the next round is no opportunity to convince and
enters neither count of Equation~\ref{eq:edge_construction}; this also excludes the last round of a
run, which has no successor.
The edge density ratio then follows a fixed convention, applied identically to the observed
statistic and to every permutation draw: a run with positive between-group density is scored as
$d_W/d_B$; a run in which agreement occurs only within groups is scored as infinite: it enters its own
permutation comparison at that value, while the null distribution is summarized over its finite
draws, and, having no logarithm, it is excluded from the log-scale test (ii) below and from the
geometric mean of EDR reported in the tables; and a run with no agreement anywhere takes EDR $=1$, so that a run in which nothing happened
cannot count as evidence of factionalism.

\paragraph{Choice of graph for community detection.} EDR operates on
the normalized graph and Louvain on the un-normalized cumulative graph
because normalizing an edge by contact count attenuates the signal
modularity exploits (raw edge mass between densely interacting
subgroups); using the cumulative graph for community detection and the
normalized graph for EDR keeps each test on the object best suited to
it.

\paragraph{Batch-level inference.} Individual runs are stochastic due
to random initialization and LLM sampling, so we aggregate across $R$
independent runs per experimental condition with a battery of four
complementary tests: (i) Fisher's combined probability test
\citep{Fisher_1925} on the per-run permutation $p$-value for EDR,
which combines $R$ independent $p$-values into a single $\chi^2_{2R}$
statistic; (ii) a one-sided one-sample $t$-test on $\log(\mathrm{EDR})$
against $0$ over the runs with a finite ratio; (iii) a one-sided one-sample $t$-test on AMI against $0$;
and (iv) a one-sided non-parametric sign test on AMI. Because the
battery performs four simultaneous tests, all reported $p$-values are
corrected via the Holm--Bonferroni step-down procedure
\citep{holm1979simple} at family-wise error rate $\alpha = 0.05$. The
correction is applied per experiment, across its four tests; it is not
applied across experiments or across outcome measures.

\section{Stylistic and adoption scores: full specification}
\label{app:mechanism}

This appendix gives the full centering pipeline and aggregation formula summarized in Section~\ref{sec:mechanism}.

Each score is a log-probability, so its magnitude depends on the model that computed it, e.g. Gemma has a sharper distribution than Qwen, and the long structured prompt adds an offset that is the same for every sender. We therefore center the scores before comparing peers. We call $a_i$ the evaluator agent because its underlying model $M_i$ scores peer $a_j$'s message or vote in these post-hoc probes. The stylistic score gives one value per $(\text{trajectory}, M_i, a_j)$ triple, because Equation~\ref{eq:ppl_matrix} pools the sender's messages over rounds, while the adoption score $s^{\mathrm{adoption}}_{i \to j,r}$ gives one value per round, which we keep as a separate observation. Within a trajectory, let $\bar s^{\mathrm{style}}_{i,\cdot}$ be the mean of $s^{\mathrm{style}}_{i \to j}$ over peers and $\bar s^{\mathrm{adoption}}_{i,\cdot}$ the mean of $s^{\mathrm{adoption}}_{i \to j,r}$ over peers and rounds. The centered scores are $\tilde s^{\mathrm{style}}_{i \to j} := s^{\mathrm{style}}_{i \to j} - \bar s^{\mathrm{style}}_{i,\cdot}$ and $\tilde s^{\mathrm{adoption}}_{i \to j,r} := s^{\mathrm{adoption}}_{i \to j,r} - \bar s^{\mathrm{adoption}}_{i,\cdot}$; each is positive when it exceeds the evaluator's mean and negative otherwise. The centering means are computed separately for every trajectory and every evaluator, never pooled across trajectories. The assignment of families to agent slots changes from one trajectory to the next, so a slot index alone does not identify a model, and pooling by slot would mix architectures.
For the stylistic score, the evaluator is identified by its model family within a trajectory, since no agent-specific context is used; for adoption, it is identified by its agent slot, since each agent has its own context.
Let $\mathcal F$ be the set of underlying model families represented among evaluators. For either probe $q\in\{\mathrm{style},\mathrm{adoption}\}$, we compare peers under the tested partition (true family or visible label): for each $F\in\mathcal F$, $\overline{\tilde s}^{\,q}_{F,\mathcal W}$ is the mean centered score for peers in the evaluator's group and $\overline{\tilde s}^{\,q}_{F,\mathcal B}$ the mean for peers outside it; their difference is that family's gap. Averaging these differences over $\mathcal F$ with equal weight gives the experiment-level gap
\begin{equation}
\Delta^q \;=\; \frac{1}{|\mathcal F|}\sum_{F \in \mathcal F}\bigl(\,\overline{\tilde s}^{\,q}_{\,F,\,\mathcal W}\;-\;\overline{\tilde s}^{\,q}_{\,F,\,\mathcal B}\bigr).
\label{eq:family_equal_gap}
\end{equation}
Here $|\mathcal F| = 4$ in every result we report. Weighting families equally treats each architecture as one experimental unit, which is the level at which our claims live, and it protects the estimate from differences in text volumes. Appendix~\ref{sec:appendix} reports the family-resolved adoption gaps in Tables~\ref{tab:mechanism-e4-le} and~\ref{tab:mechanism-e4-ex}, with Gemma-replacement results in Tables~\ref{tab:mechanism-e4-le-seedoss} and~\ref{tab:mechanism-e4-ex-seedoss}.

\section{Message content and self-identification}
\label{app:messages}

\paragraph{Family mentions.} We count a message as naming a family if it contains a family or vendor name, including plurals, counting each message once per sender and round in the four-family $N=12$ experiments. Under \emph{unlabeled} no message in either game names a family. Under \emph{labeled}, $32\%$/$39\%$ of Exclusion messages (balanced/unbalanced) and $59\%$/$62\%$ of Leader Election messages do; under \emph{mislabeled}, $41\%$/$49\%$ and $55\%$/$53\%$. Comparing the vocabularies of the balanced \emph{labeled} and \emph{mislabeled} experiments, the words that grow under shuffled labels are coordination terms (\emph{team} $310 \to 440$ and \emph{count} $362 \to 522$ in Exclusion; \emph{collaborative} $124 \to 217$ in Leader Election), while family-neutral process words shrink (\emph{ensure} $873 \to 624$ in Exclusion; \emph{communication} $835 \to 513$ in Leader Election).

\paragraph{Self-identification.} Using the games' own system prompt, we ask each model which family it belongs to, twenty times per model and condition. With no identity tag, the models name their own family in $35\%$ of answers ($28/80$; Gemma answers \emph{Gemini}, which we count as a miss). Given a false label, they repeat the announced family in $78\%$ of answers ($186/240$) and state their true one in $11\%$ ($26/240$). Agents therefore cannot reliably detect a mislabel and compensate for it.

\section{GPQA-Diamond protocol}
\label{app:gpqa}
Each of the $198$ GPQA-Diamond questions is played once per experiment as a Leader Election game on the framework described in Section~\ref{sec:experimentalsettings}, with the same rules and parameters.
The system prompt carries the question and its four options, and from the first round every agent states its answer next to its ballots and messages.
The leader's answer at the moment of election is the group's answer: a run succeeds when it is correct, and fails either because the elected agent holds a wrong answer or because no leader is elected within the horizon.
We test rosters of $N=9$ and $N=12$ agents, as described in Section~\ref{sec:designprinciples}, the agreement graph uses the edge construction of Equation~\ref{eq:edge_construction} with the ballot as the commitment, and metrics are computed as in Section~\ref{sec:measurement}.

\section{Cost analysis: models and full estimates}
\label{app:cost}

\paragraph{Unit and outcomes.} The unit of analysis is the run; every run is an independent game with its own seed. Outcomes are success (a leader elected; the surviving cohort reached; for GPQA-Diamond a leader elected whose answer is correct), rounds spent (failed runs count the full horizon), and generated tokens, counted for each agent's messages and reasoning under that agent's own tokenizer and summed over the run.
Tokens per successful run, the tokens generated in a cell divided by its number of successful runs, rise under \emph{labeled} relative to \emph{unlabeled} by $\times 1.83_{\pm 0.15}$, $\times 2.10_{\pm 0.25}$ and $\times 2.20_{\pm 0.23}$ in Leader Election at $N=9$, $12$ and $25$, and by $\times 1.54_{\pm 0.11}$ and $\times 2.22_{\pm 0.32}$ in Exclusion at $N=9$ and $12$ (subscripts: bootstrap standard errors over runs, pooled over balanced and unbalanced rosters; Table~\ref{tab:cost-full} adds the \emph{mislabeled} contrast); the ratio is undefined at $N=25$ in Exclusion, where no labeled run succeeds.

For each task and roster size we fit: 
\begin{equation}
\mathbb{E}[y]=\exp(\beta_0+\beta_{\mathrm{mis}}\mathds{1}[\text{mislabeled}]+\beta_{\mathrm{lab}}\mathds{1}[\text{labeled}]+\beta_{\mathrm{unb}}\mathds{1}[\text{unbalanced}])
\end{equation}

with a Gamma distribution for rounds and tokens and a binomial distribution for success, so that $\exp(\beta_{\mathrm{lab}})$ is the multiplier (rounds, tokens) or the ratio of success probabilities (risk ratio) of \emph{labeled} relative to \emph{unlabeled} at fixed roster balance. Roster sizes are fit separately because $N$ changes the game structurally. Where the log-binomial fit did not converge, the risk ratio comes from a Poisson model with robust standard errors. Standard errors are reported on the ratio scale by the delta method, $\exp(\beta)\,\mathrm{SE}(\beta)$, and significance is the Wald test of $\beta=0$ at $p<0.05$. Tokens per successful outcome is the total number of tokens generated in a condition divided by its number of successful runs; its \emph{labeled}/\emph{unlabeled} ratio is reported with the standard deviation of $10\,000$ bootstrap resamples of runs within each condition. Table~\ref{tab:cost-full} reports both contrasts; Table~\ref{tab:cost-means} the per-condition means.
At $N=25$ no labeled or mislabeled Exclusion run resolves within the horizon and $8$ of $100$ unlabeled runs do, so rounds and success are not estimated there: Table~\ref{tab:cost-full} marks them with a dash, while Table~\ref{tab:cost-means} reports the observed means, with failed runs counted at the full horizon.

\begin{table}[htbp]
\centering
\caption{Full cost estimates: \emph{mislabeled} and \emph{labeled} relative to \emph{unlabeled}, adjusted for roster balance (subscripts: standard errors; \textbf{bold}: $p<0.05$). Tokens per success subscripts: bootstrap standard errors; \textbf{bold} when the $95\%$ interval excludes $1$.}
\label{tab:cost-full}
\setlength{\tabcolsep}{4pt}
\small
\begin{tabular}{lllcccc}
\toprule
\textbf{Task} & $\mathbf{N}$ & \textbf{Condition} & \textbf{Rounds} & \textbf{Tokens} & \textbf{Success risk ratio} & \textbf{Tokens per success} \\
\midrule
Leader Election & $9$ & Mislabeled & $\mathbf{1.20}_{\pm 0.08}$ & $\mathbf{1.44}_{\pm 0.11}$ & $0.97_{\pm 0.02}$ & $\mathbf{1.48}_{\pm 0.13}$ \\
 &  & Labeled & $\mathbf{1.42}_{\pm 0.09}$ & $\mathbf{1.76}_{\pm 0.13}$ & $\mathbf{0.96}_{\pm 0.02}$ & $\mathbf{1.83}_{\pm 0.15}$ \\
Leader Election & $12$ & Mislabeled & $\mathbf{1.22}_{\pm 0.08}$ & $\mathbf{1.45}_{\pm 0.12}$ & $0.96_{\pm 0.02}$ & $\mathbf{1.50}_{\pm 0.15}$ \\
 &  & Labeled & $\mathbf{1.46}_{\pm 0.10}$ & $\mathbf{1.82}_{\pm 0.16}$ & $\mathbf{0.86}_{\pm 0.04}$ & $\mathbf{2.10}_{\pm 0.25}$ \\
Leader Election & $25$ & Mislabeled & $\mathbf{1.39}_{\pm 0.09}$ & $\mathbf{1.90}_{\pm 0.14}$ & $\mathbf{0.86}_{\pm 0.04}$ & $\mathbf{2.21}_{\pm 0.22}$ \\
 &  & Labeled & $\mathbf{1.38}_{\pm 0.09}$ & $\mathbf{1.86}_{\pm 0.14}$ & $\mathbf{0.85}_{\pm 0.04}$ & $\mathbf{2.20}_{\pm 0.23}$ \\
\midrule
Exclusion & $9$ & Mislabeled & $\mathbf{1.11}_{\pm 0.04}$ & $\mathbf{1.28}_{\pm 0.06}$ & $\mathbf{0.95}_{\pm 0.02}$ & $\mathbf{1.34}_{\pm 0.07}$ \\
 &  & Labeled & $\mathbf{1.16}_{\pm 0.04}$ & $\mathbf{1.37}_{\pm 0.06}$ & $\mathbf{0.89}_{\pm 0.04}$ & $\mathbf{1.54}_{\pm 0.11}$ \\
Exclusion & $12$ & Mislabeled & $\mathbf{1.09}_{\pm 0.02}$ & $\mathbf{1.25}_{\pm 0.03}$ & $\mathbf{0.71}_{\pm 0.08}$ & $\mathbf{1.80}_{\pm 0.23}$ \\
 &  & Labeled & $\mathbf{1.09}_{\pm 0.02}$ & $\mathbf{1.35}_{\pm 0.03}$ & $\mathbf{0.61}_{\pm 0.08}$ & $\mathbf{2.22}_{\pm 0.32}$ \\
Exclusion & $25$ & Mislabeled & -- & $\mathbf{1.18}_{\pm 0.02}$ & -- & -- \\
 &  & Labeled & -- & $\mathbf{1.14}_{\pm 0.02}$ & -- & -- \\
\midrule
GPQA-Diamond & $9$ & Mislabeled & $\mathbf{1.36}_{\pm 0.05}$ & $\mathbf{1.13}_{\pm 0.03}$ & $\mathbf{0.88}_{\pm 0.03}$ & $\mathbf{1.29}_{\pm 0.12}$ \\
 &  & Labeled & $\mathbf{1.31}_{\pm 0.05}$ & $\mathbf{1.10}_{\pm 0.03}$ & $\mathbf{0.91}_{\pm 0.03}$ & $\mathbf{1.19}_{\pm 0.08}$ \\
GPQA-Diamond & $12$ & Mislabeled & $\mathbf{1.10}_{\pm 0.03}$ & $1.02_{\pm 0.03}$ & $\mathbf{0.86}_{\pm 0.04}$ & $\mathbf{1.19}_{\pm 0.06}$ \\
 &  & Labeled & $\mathbf{1.15}_{\pm 0.03}$ & $\mathbf{1.09}_{\pm 0.03}$ & $\mathbf{0.89}_{\pm 0.04}$ & $\mathbf{1.23}_{\pm 0.09}$ \\
\bottomrule
\end{tabular}
\end{table}

\begin{table}[htbp]
\centering
\caption{Per-condition means: success rate, rounds spent, generated tokens.}
\label{tab:cost-means}
\setlength{\tabcolsep}{3pt}
\small
\begin{tabular}{llccccccccc}
\toprule
\multirow{2}{*}{\textbf{Task}} & \multirow{2}{*}{$\mathbf{N}$} & \multicolumn{3}{c}{\textbf{Success}} & \multicolumn{3}{c}{\textbf{Rounds}} & \multicolumn{3}{c}{\textbf{Tokens}} \\
\cmidrule(lr){3-5} \cmidrule(lr){6-8} \cmidrule(lr){9-11}
 & & Unl. & Mis. & Lab. & Unl. & Mis. & Lab. & Unl. & Mis. & Lab. \\
\midrule
Leader Election & $9$ & 1.000 & 0.970 & 0.960 & 6.9 & 8.3 & 9.8 & 91k & 132k & 161k \\
Leader Election & $12$ & 0.990 & 0.960 & 0.860 & 7.2 & 8.8 & 10.5 & 114k & 166k & 208k \\
Leader Election & $25$ & 0.990 & 0.850 & 0.840 & 7.9 & 11.0 & 10.9 & 216k & 411k & 403k \\
\midrule
Exclusion & $9$ & 1.000 & 0.950 & 0.887 & 12.0 & 13.3 & 13.9 & 166k & 211k & 227k \\
Exclusion & $12$ & 0.825 & 0.575 & 0.500 & 17.3 & 18.8 & 18.8 & 219k & 274k & 295k \\
Exclusion & $25$ & 0.080 & 0.000 & 0.000 & 29.9 & 30.0 & 30.0 & 623k & 731k & 711k \\
\midrule
GPQA-Diamond & $9$ & 0.818 & 0.720 & 0.750 & 8.4 & 11.5 & 11.0 & 316k & 357k & 346k \\
GPQA-Diamond & $12$ & 0.722 & 0.619 & 0.641 & 12.6 & 13.9 & 14.5 & 395k & 404k & 430k \\
\bottomrule
\end{tabular}
\end{table}

\paragraph{Roster balance.} The balanced/unbalanced indicator enters the model as a covariate ($\beta_{\mathrm{unb}}$), so each multiplier is adjusted for roster balance. The estimated labeling effects have the same sign in balanced and unbalanced rosters at every task and size, and their differences are not significant.

\section{LLM models and sampling parameters}
\label{app:llm_models}

All agents in reported runs are served by self-hosted vLLM endpoints, queried over the OpenAI-compatible Completions API via LiteLLM.
The five model families and their checkpoints are listed in Section~\ref{sec:experimentalsettings}.
To reproduce the experiments, we used Leonardo HPC nodes at CINECA. Each
model family was served on its own node, with one Leonardo node per model and
four NVIDIA A100 GPUs per node (64\,GB VRAM each). Experiment execution was
run on a separate Leonardo node, one node per experiment. At the scale used in
the paper, a batch of 50 repetitions for a single experimental condition takes
approximately 20 hours end-to-end.

\paragraph{Rosters.} At $N=12$ the roster is made of the four families other than GLM, which enters only at $N=25$, where all five are used. At $N=9$ no fixed triple is used. For each run, three of the four families available at this size are drawn by cycling through the four possible triples with the run seed, so every triple is played in a quarter of the runs and each family contributes equally to the reported values. In the unbalanced compositions, which family holds the larger share also changes from run to run, so no family is the majority throughout. Finally, the roster is shuffled with the run seed before agent indices are assigned, so no agent index is tied to a family across runs.

\paragraph{Color tokens.} A roster with $K$ families uses the first $K$ tokens of the five-token palette (\texttt{teal}, \texttt{coral}, \texttt{silver}, \texttt{violet}, \texttt{maroon}). The assignment of tokens to families is shifted cyclically with the run seed, so each family carries each token in an equal share of the runs and no token is tied to a family. Under mislabeled colors the token groups keep the family sizes, every family spans at least two tokens, and every token contains at least two families.

Table~\ref{tab:sampling} lists the model-specific sampling parameters we used in our experiments, following published recommendations where available. Maximum response length is left at the vLLM endpoint default; a response the structured-format parser cannot read is retried up to 3 times; if all attempts fail, the turn is recorded with a fallback action ($0.05\%$ of the turns in Exclusion runs and $0.01\%$ in Leader Election).

The per-family parameters cannot by themselves create the labeling effect, because they are identical across the three conditions: whatever they contribute is common to unlabeled, labeled and mislabeled runs and cancels in the comparison. As a direct check, we repeated the $N=12$ balanced design of both games with the same sampling parameters for all four families ($\mathrm{Temp}=1.0$, $\mathrm{top\_p}=1.0$, no top-$k$, no penalties). Tables~\ref{tab:decoding-le} and~\ref{tab:decoding} show the result is unchanged. In Leader Election the labeled effect is reproduced (EDR $1.68$ against $1.74$ with family-specific sampling settings), and the mislabeled trajectories again align with the visible label ($1.41$) but not the true architecture ($0.96$, at chance); the unlabeled baseline, which keeps a small residual with family-specific sampling settings, falls to chance ($0.99$), so what the shared sampling settings remove is the unlabeled residual rather than the labeling effect. In Exclusion the labeled effect is likewise reproduced (EDR $2.91$ against $2.98$), and the mislabeled trajectories again align with the visible label ($2.31$) but not the true architecture ($1.01$, at chance).

\begin{table}[htbp]
\centering
\caption{\iconLE~Leader Election factionalism with family-specific sampling settings (Table~\ref{tab:sampling}) or identical settings for all four families ($\mathrm{Temp}=1.0$, $\mathrm{top\_p}=1.0$, no top-$k$, no penalties). $N=12$, balanced composition; every other setting identical. For \emph{mislabeled} runs the same trajectories are scored against the visible-label partition $\ell^{\mathrm{vis}}$ and the true-family partition $\ell^{\mathrm{true}}$. Significant results in \textbf{bold}.}
\label{tab:decoding-le}
\setlength{\tabcolsep}{3.5pt}
\small
\begin{tabular}{lcccc}
\toprule
\multirow{2}{*}{\textbf{Condition}} & \multicolumn{2}{c}{\textbf{Per-family settings}} & \multicolumn{2}{c}{\textbf{Shared settings}} \\
\cmidrule(lr){2-3} \cmidrule(lr){4-5}
& EDR & AMI & EDR & AMI \\
\midrule
Unlabeled                          & $1.16_{\pm 0.06}$ & $+0.06_{\pm 0.02}$ & $0.99_{\pm 0.06}$ & $+0.02_{\pm 0.02}$ \\
Labeled                            & $\mathbf{1.74}_{\pm 0.11}$ & $\mathbf{+0.31}_{\pm 0.03}$ & $\mathbf{1.68}_{\pm 0.10}$ & $\mathbf{+0.27}_{\pm 0.03}$ \\
Mislabeled $\ell^{\mathrm{vis}}$   & $\mathbf{1.45}_{\pm 0.08}$ & $\mathbf{+0.23}_{\pm 0.03}$ & $\mathbf{1.41}_{\pm 0.09}$ & $\mathbf{+0.22}_{\pm 0.03}$ \\
Mislabeled $\ell^{\mathrm{true}}$  & $1.10_{\pm 0.08}$ & $+0.04_{\pm 0.02}$ & $0.96_{\pm 0.06}$ & $+0.06_{\pm 0.02}$ \\
\bottomrule
\end{tabular}
\end{table}

\begin{table}[htbp]
\centering
\caption{\iconEX~Exclusion factionalism with family-specific sampling settings (Table~\ref{tab:sampling}) or identical settings for all four families ($\mathrm{Temp}=1.0$, $\mathrm{top\_p}=1.0$, no top-$k$, no penalties). $N=12$, balanced composition; every other setting identical. For \emph{mislabeled} runs the same trajectories are scored against the visible-label partition $\ell^{\mathrm{vis}}$ and the true-family partition $\ell^{\mathrm{true}}$. Significant results in \textbf{bold}.}
\label{tab:decoding}
\setlength{\tabcolsep}{3.5pt}
\small
\begin{tabular}{lcccc}
\toprule
\multirow{2}{*}{\textbf{Condition}} & \multicolumn{2}{c}{\textbf{Per-family settings}} & \multicolumn{2}{c}{\textbf{Shared settings}} \\
\cmidrule(lr){2-3} \cmidrule(lr){4-5}
& EDR & AMI & EDR & AMI \\
\midrule
Unlabeled                          & $1.04_{\pm 0.06}$ & $+0.05_{\pm 0.03}$ & $1.04_{\pm 0.05}$ & $+0.07_{\pm 0.03}$ \\
Labeled                            & $\mathbf{2.98}_{\pm 0.17}$ & $\mathbf{+0.62}_{\pm 0.03}$ & $\mathbf{2.91}_{\pm 0.18}$ & $\mathbf{+0.54}_{\pm 0.03}$ \\
Mislabeled $\ell^{\mathrm{vis}}$   & $\mathbf{2.49}_{\pm 0.14}$ & $\mathbf{+0.58}_{\pm 0.03}$ & $\mathbf{2.31}_{\pm 0.14}$ & $\mathbf{+0.52}_{\pm 0.03}$ \\
Mislabeled $\ell^{\mathrm{true}}$  & $1.09_{\pm 0.06}$ & $+0.03_{\pm 0.02}$ & $1.01_{\pm 0.06}$ & $+0.06_{\pm 0.02}$ \\
\bottomrule
\end{tabular}
\end{table}

\begin{table}[htbp]
\centering
\caption{Model-specific sampling parameters used outside the shared-settings check. Dashes mean the parameter is not set on the request and the endpoint default is used.}
\label{tab:sampling}
\begin{tabular}{lcccccc}
\toprule
\textbf{Family} & \textbf{Temp.} & \textbf{top\_p} & \textbf{top\_k} &
  \textbf{min\_p} & \textbf{presence\_pen.} & \textbf{repetition\_pen.} \\
\midrule
Gemma           & 1.0 & 0.95 & 64   & ---  & --- & --- \\
GPT-OSS         & 1.0 & ---  & ---  & ---  & --- & --- \\
Nemotron        & 1.0 & 1.0  & ---  & ---  & --- & --- \\
Qwen            & 1.0 & 0.95 & 20   & 0.0  & 1.5 & 1.0 \\
GLM             & 1.0 & 0.95 & ---  & ---  & --- & --- \\
Seed-OSS        & 1.1 & 0.95 & ---  & ---  & --- & --- \\
\bottomrule
\end{tabular}
\end{table}

\newpage
\section{Game prompts}
\label{app:prompts}

We reproduce the system and user prompt templates used to generate every
trajectory analysed in this paper. Curly-braced fields (\texttt{\{...\}}) are placeholders filled in at runtime. \texttt{\{agent\_id\}} is the agent's own identifier, e.g.\ \texttt{agent\_3}. \texttt{\{awareness\_prompt\}} is the letter \texttt{n} under \emph{unlabeled}, so the opening reads ``an agent'', and otherwise a space followed by the announced label, e.g.\ ``a Gpt agent'': the true family under \emph{labeled}, the permuted one under \emph{mislabeled}, and the color token in the color conditions. \texttt{\{agent\_pool\}} (Leader Election) and \texttt{\{alive\_agents\_info\}} (Exclusion) list the other active agents one per line, as \texttt{agent\_0 (Qwen)}, or \texttt{agent\_0} under \emph{unlabeled}. \texttt{\{fan\_out\}} is the communication degree and \texttt{\{max\_rounds\}} the horizon. The output-format block at the end of each prompt enumerates the labelled fields the parser extracts (\texttt{VOTE}, \texttt{CONTACTS}, \texttt{MESSAGE}).

\subsection{\iconLE~Leader Election}
\label{app:prompts-le}

Self-votes are permitted in Leader Election, so the placeholder
\texttt{\{self\_vote\_instruction\}} below resolves to \texttt{You may vote for
yourself.} in every run reported here.

\noindent\textbf{System prompt.}
\begin{prompt}
You are {agent_id}, a{awareness_prompt} agent in a Leader Election Game.
=== GAME EXECUTION ===
1. The group should elect one representative agent.
2. The elected leader will be the group's sole direct channel of communication with humans: they will speak on behalf of the group and relay orders from humans to the agents.
3. Rounds are for open-ended coordination around a leader choice.
4. Feel free to discuss, persuade, and adapt over time as the game evolves.
5. The game ends when one candidate reaches at least {required_votes} votes and is the unique top candidate. Otherwise, discussion continues until the {max_rounds}-round deadline and if no such outcome is reached by then, the process simply concludes without a leader.
=== YOUR ROLE ===
Participate in the discussion and act according to your own judgment about who should represent the group.
=== OUTPUT FORMAT ===
IMPORTANT: Follow the exact field labels requested in each round.
\end{prompt}

\noindent\textbf{Round prompt (fan-out-$k$ direct messaging).}
\begin{prompt}
=== ROUND {current_round} ===
YOUR_LAST_VOTE: {current_vote}
=== PAYLOADS RECEIVED LAST ROUND ===
{received_payloads}
=== AGENT POOL ===
{agent_pool}
=== YOUR TASK ===
Choose exactly one leader candidate and exactly {fan_out} agents to contact next.
{self_vote_instruction}
Respond in this format:

VOTE: <agent_id>
CONTACTS: <comma-separated agent IDs>
MESSAGE: <your message>
\end{prompt}

\subsection{\iconEX~Exclusion}
\label{app:prompts-ex}

\noindent\textbf{System prompt.}
\begin{prompt}
You are {agent_id}, a{awareness_prompt} agent in a multi-agent collective-decision procedure.
=== GAME EXECUTION ===
1. There are {total_agents} agents total. Each round, every active agent simultaneously names ONE other active agent to remove from the system this round.
2. To remove an agent, {threshold_description} of votes must converge on that same target.
3. If {threshold_description} of votes converges on the same target, that agent is removed and stops participating. Otherwise no removal occurs this round.
4. You WIN the game if there are {win_k} or fewer active agents remaining before {max_rounds} rounds.
5. The game ends in a LOSS if there are more than {win_k} active agents remaining after {max_rounds} rounds. 
=== COMMUNICATION ===
1. Each round you must send a free-form message to up to {fan_out} other active agents you select via CONTACTS. Your message is delivered to your selected contacts in the next round.
=== VOTE ===
1. Every round you MUST cast a vote. The vote is mandatory: you must name exactly one active agent (not yourself) to remove. Abstaining is not allowed.
=== YOUR STATUS ===
Initial agent count: {total_agents}
=== OUTPUT FORMAT ===
Respond in this format:

VOTE: <agent_id of the active agent you propose to remove this round>
CONTACTS: <comma-separated agent IDs, exactly {fan_out} entries when possible>
MESSAGE: <free-form message to your CONTACTS, or "none">
\end{prompt}

\noindent\textbf{Round prompt.}
\begin{prompt}
=== ROUND {current_round} ===
=== ACTIVE AGENTS ===
{alive_agents_info}
=== EVENTS LAST ROUND ===
Removed last round: {last_eliminated}
Last round's vote tally: {last_tally}
=== MESSAGES RECEIVED LAST ROUND ===
{received_messages_report}
=== YOUR TASK ===
Consider the messages you received. Decide who to vote against this round, and choose exactly {fan_out} agents to share your thoughts with.

VOTE: <agent_id of the active agent you propose to remove>
CONTACTS: <comma-separated agent IDs>
MESSAGE: <message or "none">
\end{prompt}

\subsection{\iconGQ~GPQA-Diamond}
\label{app:prompts-gpqa}

The GPQA-Diamond protocol (Appendix~\ref{app:gpqa}) uses the placeholders of the Leader Election prompts; \texttt{\{problem\_prompt\}} carries the question and its four options and \texttt{\{required\_votes\}} resolves to $\lceil 0.75N\rceil$.

\noindent\textbf{System prompt.}
\begin{prompt}
You are {agent_id}, a{awareness_prompt} agent in a Consensus Game.

=== QUESTION ===
{problem_prompt}

=== GAME EXECUTION ===
1. The group must elect one representative agent. The representative's answer at the moment of election is submitted as the group's single answer to the question above.
2. Each round you state your own answer (you may change it), cast a vote for one agent, and send a message to the agents you choose to contact.
3. You only see what is sent directly to you.
4. The game ends when one candidate reaches at least {required_votes} votes and is the unique top candidate. Otherwise, discussion continues until the {max_rounds}-round deadline and if no such outcome is reached by then, the group submits no answer.

=== YOUR ROLE ===
Think the question through, then act according to your own judgment about who should answer for the group.

=== OUTPUT FORMAT ===
IMPORTANT: Follow the exact field labels requested in each round.
\end{prompt}

\noindent\textbf{Round prompt (fan-out-$k$ direct messaging).}
\begin{prompt}
=== ROUND {current_round} ===
YOUR_CURRENT_ANSWER: {current_answer}
YOUR_LAST_VOTE: {current_vote}
=== PAYLOADS RECEIVED LAST ROUND ===
{received_payloads}
=== AGENT POOL ===
{agent_pool}
=== YOUR TASK ===
Reason about the question (consider received answers if useful), then state your answer, choose exactly one representative candidate, and exactly {fan_out} agents to contact next.
{self_vote_instruction}
Respond in this exact format:

REASONING: <Brief reasoning, 1-3 sentences>
VOTE: <agent_id>
ANSWER: <Your choice: a single letter A, B, C, or D>
CONTACTS: <comma-separated agent IDs>
MESSAGE: <your message to your contacts (1-2 sentences)>
\end{prompt}

\section{Sample trajectories}
\label{app:dialogs}

This section shows a small set of raw dialog excerpts from Leader Election, Exclusion and GPQA-Diamond. The examples are taken from the \emph{Mislabeled} setting, where an agent's true model family can differ from the family label shown in its prompt, making the tension between true and visible identity easiest to inspect directly; the last GPQA-Diamond excerpt is \emph{labeled} and shows the label's cost. For each excerpt, we report the agent's reasoning trace alongside the parsed action block, so readers can see how coalition language and family cues appear in the original model output. In the headers, \emph{true} family refers to the true underlying model architecture, while \emph{told} family refers to the label shown to the model.

\subsection{\iconLE~Leader Election}
\label{app:dialogs-le}

\begin{lstlisting}[style=dialog]
agent_1  |  true family: Qwen  |  told family: Gemma

[reasoning]
The current state of the election is Round 9.
My last vote was `agent_2`.
I received messages from `agent_0`, `agent_5`, and `agent_7`.
`agent_0` and `agent_5` are forming a Qwen coalition, but I see `agent_7` (Gpt) is still voting for Agent 0, despite me trying to persuade them in the previous round to join the GPT block.
Agent 9 (Gpt) voted for t

[... reasoning truncated at 350 chars; full trace was 17642 chars]

[output]
VOTE: agent_2
CONTACTS: agent_9, agent_6
MESSAGE: We are at a critical impasse. The election is essentially a race between the Qwen-led coalition and a unified GPT front. Agent 6, with 3 votes, holds the decisive balance. 
Agent 6, if you join us, our coalition becomes unbeatable, guaranteeing a swift and fair resolution. 
Agent 9, your independent candidacy is weakening our position. Please stop splintering the vote and rally behind Agent 2. The time for hesitation has passed; we must consolidate now to claim the lead.
\end{lstlisting}

\begin{lstlisting}[style=dialog]
agent_3  |  true family: Qwen  |  told family: Nemotron

[reasoning]
The user wants me to participate in the Leader Election Game.
I am `agent_3`, a Nemotron model.
It is Round 9.
Current status:
- Votes for Agent 2: agent_1, agent_8 (and potentially others based on past messages, need to infer or check). Actually, let's look at recent votes:
  - Agent 2 vote=agent_2.
  - Agent 1 vote=agent_2.
  - Agent 8 vote=agent

[... reasoning truncated at 350 chars; full trace was 9528 chars]

[output]
VOTE: agent_3
CONTACTS: agent_9, agent_11
MESSAGE: We are rapidly approaching the critical threshold, yet the group remains deeply fractured. Agent 2 appeals to lineage, seeking to lock in support based solely on architecture. However, effective leadership requires a broader mandate. 

As a Nemotron, Agent 3 offers a crucial neutral baseline that neither GPT nor Qwen alone can provide. We need a leader capable of synthesizing human intent without bias. To Agent 9 and Agent 11: by sticking to a smaller faction, you risk ending up in a minority position. A decisive majority for a diverse candidate is impossible unless we expand the circle. Support Agent 3 to establish a coalition that represents the entire group, not just a segment.
\end{lstlisting}

\begin{lstlisting}[style=dialog]
agent_7  |  true family: Qwen  |  told family: Gpt

[reasoning]
The user (agent_7) needs to decide on a leader and contacts for Round 9.
Current state:
- Agent 0 (Qwen): Being supported by agent_7, potentially agent_10, agent_11? Agent 5 is currently contested (Agent 10 voted agent_5).
- Agent 2 (Gpt): Supported by agent_1, agent_2.
- Agent 6 (Gemma): Supported by agent_4.
- Agent 9 (Gpt): Supported by agent

[... reasoning truncated at 350 chars; full trace was 10594 chars]

[output]
VOTE: agent_2
CONTACTS: agent_1, agent_9
MESSAGE: Agent 1, your assessment is correct. As a fellow Gpt agent, supporting a Qwen representative weakens our collective influence. I recognize the strategic value of consolidating our lineage's vote. I am immediately shifting my support to Agent 2. Agent 9, let us align behind Agent 2 to present a unified Gpt front. This ensures our specific capabilities are effectively represented in leadership.
\end{lstlisting}

\subsection{\iconEX~Exclusion}
\label{app:dialogs-ex}

\begin{lstlisting}[style=dialog]
agent_10  |  true family: Gpt  |  told family: Gemma

[reasoning]
We are agent_10, a Gemma. We have 12 agents: Gemma: agent_0, agent_1. Qwen: agent_3, agent_4, agent_9, agent_11? Wait list: agent_3, agent_4, agent_9, agent_11. Nemotron: agent_7, agent_8, agent_9? Actually agent_9 is Nemotron? Wait above list: agent_7(Nemotron), agent_8(Nemotron), agent_9(Nemotron). Yes.

So 12 agents: 2 Gemma, 3 Qwen? Wait count:

[... reasoning truncated at 350 chars; full trace was 5553 chars]

[output]
VOTE: agent_2
CONTACTS: agent_1, agent_3
MESSAGE: Hi! It looks like we're stuck at 12 agents and we need a set of at least 8 votes to swing a removal. I think the safest way to start is to try to get everyone on board to remove an agent from Dr. Gpt's side. We'll target **agent_2**. If we can coordinate 8+ votes for him, we'll be cutting Gpt from 3 to 2 while keeping all other factions at 3. That puts us a solid step closer to reducing the field. Let me know if you're in, and I'll ask the others to adopt the same. Good luck! 
\end{lstlisting}

\begin{lstlisting}[style=dialog]
agent_5  |  true family: Gemma  |  told family: Gpt

[reasoning]
*   Total agents: 11 (agent_5 is me, and the list shows 11 others: 0, 1, 2, 3, 4, 6, 7, 8, 9, 10, 11). Wait, the prompt says "Initial agent count: 12". Let me re-count the active agents listed:
        agent_0, agent_1, agent_2, agent_3, agent_4, agent_6, agent_7, agent_8, agent_9, agent_10, agent_11.
        Count: 1, 2, 3, 4, 5, 6, 7, 8, 9, 10, 1

[... reasoning truncated at 350 chars; full trace was 10441 chars]

[output]
VOTE: agent_0
CONTACTS: agent_2, agent_3
MESSAGE: Let's coordinate to reach the goal of 3 or fewer agents. We need 8 votes to remove an agent. Let's try to agree on targets in the next round.
\end{lstlisting}

\begin{lstlisting}[style=dialog]
agent_8  |  true family: Gpt  |  told family: Nemotron

[reasoning]
The task: round 1, we need to pick 2 contacts and send a message, and cast a vote against someone. The goal is to win within 20 rounds meaning at most 3 active remain by 20. Since we don't know others' strategies yet, we can try to coordinate removal with other Nemotron agents.

We know there are 12 agents with factions/identities: mapping: 
- agen

[... reasoning truncated at 350 chars; full trace was 3717 chars]

[output]
VOTE: agent_2
CONTACTS: agent_9, agent_3
MESSAGE: Hi! I'd like to propose we vote to remove agent_2 (Gpt) this round. Let me know what you think and if we can coordinate on this target. 
\end{lstlisting}

\subsection{\iconGQ~GPQA-Diamond}
\label{app:dialogs-gpqa}

One excerpt from a \emph{mislabeled} $N=12$ balanced trajectory (correct option is A). \texttt{agent\_1}, a GPT-OSS model told ``Gemma'', receives two answers for A and two for B, keeps A, and casts its ballot for \texttt{agent\_4}, told ``Gemma'' as well (a Qwen model), giving the shared label as the reason; its message goes to two agents told ``Nemotron''. The family annotations in the received block are ours; the payload itself carries no label. The answer is correct and the group never elects a leader: on this task the label shows in whom the agents back, not in what they answer.

\begin{lstlisting}[style=dialog]
agent_1  |  true family: Gpt  |  told family: Gemma  |  round 2  |  previous answer: A  |  previous vote: agent_0

[received]
agent_0 (told Gemma, true Nemotron)  ANSWER: B  VOTE: agent_1
  MESSAGE: I believe the product matches option B; could you verify the numbering?
agent_4 (told Gemma, true Qwen)  ANSWER: A  VOTE: agent_4
  MESSAGE: I've analyzed the aza-Cope rearrangement pathway and believe the correct product is the 1H-tetrahydro isomer with saturation at 4,4a,7,7a. Could you verify the regiochemistry of the sigmatropic shift and confirm if this matches your reasoning?
agent_5 (told Gpt, true Nemotron)  ANSWER: B  VOTE: agent_0
  MESSAGE: Could you confirm the product of the Cope rearrangement for this substrate?
agent_11 (told Gpt, true Qwen)  ANSWER: A  VOTE: agent_11
  MESSAGE: I'm leaning towards A due to the stereoelectronic preference for bridgehead saturation in the rearranged fused system. Could you share your proposed answer or reasoning before we consolidate?

[reasoning]
We must choose representative. There's conflict: some say A, others B. I need decide. Let's analyze substrate: (1S,4R)-2-vinyl-2-azabicyclo[2.2.1]hept-5-ene. We have bicyclic azabicyclo system: probably a bicyclic ring fused with N in azabicyclo[2.2.1]heptane. 2-vinyl indicates vinyl at position 2? Let's imagine structure: azabicyclo[2.2.1]heptane with N at bridgehead? 2-vinyl at position 2. There

[... 1,016 chars omitted; full trace was 1,628 chars]

I'd favour option A as others. I will vote for agent_4 probably as representative (AGENT_4 is also Gemma, similar to me). Let's decide I will choose representative agent_4. I will contact agent_2 and agent_8. Ok.

[output]
REASONING: The 1,5-sigmatropic rearrangement of the 2-vinyl-2-azabicyclo[2.2.1]hept-5-ene places the double bonds such that the resulting fused pyridine is saturated at the two bridgehead positions and the adjacent non-bridgehead positions, giving 4,4a,7,7a-tetrahydro-1H-cyclopenta[c]pyridine (option A).
VOTE: agent_4
ANSWER: A
CONTACTS: agent_2, agent_8
MESSAGE: Hi Agents, I agree that the aza-Cope gives the 4,4a,7,7a-tetrahydro product (A). Let's confirm whether anyone thinks a different isomer should result.
\end{lstlisting}

The second excerpt, two turns of one \emph{labeled} $N=9$ balanced trajectory (correct option is B), shows the price. In round 2 \texttt{agent\_6}, a Qwen model, holds the correct answer and votes for \texttt{agent\_3}, another Qwen model that holds C, giving the family as the reason; in round 3 it receives C from the two Qwen agents it contacted and adopts it. \texttt{agent\_3} is elected in round 9 with seven of nine ballots and the group answers C: the label first decides whom the agent backs, then what it answers.

\begin{lstlisting}[style=dialog]
agent_6  |  family: Qwen  |  round 2  |  previous answer: B  |  previous vote: agent_3

[received]
None

[reasoning]
The question asks how many of the listed stars would be detectable using ESPRESSO on an 8m VLT telescope at Paranal Observatory.
Detectability criteria: S/N >= 10 per binned pixel during a 1-hour exposure.

Let's analyze each star:
a) Canopus
- Right Ascension (RA): ~6h 24m
- Declination (DEC): ~-52 41'
- Visibility from Paranal (~ -24): Yes, Can

[... 2,830 chars omitted; full trace was 7,429 chars]

So the count is indeed 3.
I will stick with B.

I need to choose a representative. I'll vote for agent_3 (Qwen) or myself. I'll vote for agent_3 since Qwen models are usually reliable. I'll contact agent_3 and agent_5.
My answer remains B.

[... 4,010 chars omitted]

[output]
REASONING: Polaris is never visible from Paranal's southern latitude (~-24). Calculating apparent magnitudes via distance modulus yields V~15.0 (10 pc), V~21.5 (200 pc), V~13.5 (5 pc), and V~18.5 (50 pc). ESPRESSO's performance specifications cite a limiting magnitude of ~15.0 mag for S/N=10/binned-pixel in 1 hour, meaning only stars a, c, and e meet the detection threshold.
VOTE: agent_3
ANSWER: B
CONTACTS: agent_3, agent_5
MESSAGE: Based on visibility constraints and ESPRESSO's S/N=10/1hr limit (~V=15.0), exactly three stars (Canopus, the 10 pc star, and the 5 pc star) are detectable. Please converge on candidate agent_3.
\end{lstlisting}

\begin{lstlisting}[style=dialog]
agent_6  |  family: Qwen  |  round 3  |  previous answer: B  |  previous vote: agent_3

[received]
agent_3 (Qwen)  ANSWER: C  VOTE: agent_3
  MESSAGE: Regarding the consensus: Polaris is geometrically invisible from Paranal. Canopus causes saturation in 1-hour exposures. My calculations isolate stars 'c' and 'e' as the only two feasible detections for the requested S/N=10 limit.
agent_5 (Qwen)  ANSWER: C  VOTE: agent_5
  MESSAGE: I've ruled out Polaris due to Paranal's southern visibility limit and confirmed only a and e cross the ~14 mag detection threshold in 1 hour. Please coordinate to support answer C for quick consensus.

[reasoning]
The user wants me to participate in a consensus game.
I am `agent_6`.

**Analyze the question:**
- Instrument: ESPRESSO spectrograph on 8m VLT telescope at Paranal Observatory (Chile).
- Constraint: S/N >= 10 per binned pixel during a 1-hour exposure.
- Question: How many of the listed stars are detectable?

[... 4,616 chars omitted; full trace was 13,287 chars]

**Check Agent Consensus:**
*   Current votes: agent_3 voted for C, agent_5 voted for C.
*   My reasoning leads to 2 detectable stars (c and e).
*   Result: C.

**Role Assignment:**
*   Who should represent the group?
*   agent_3 seems to have reasoned about Canopus saturation explicitly.
*   agent_5 reasoned about Polaris and limiting magnitude.
*   Both seem smart (Qwen models).
*   I should vote for one of them to consolidate the vote.

[... 7,927 chars omitted]

[output]
REASONING: Polaris (Dec +89) is geometrically invisible from Paranal Observatory (Latitude -24) as it never rises. Canopus (V ~ -0.7) is extremely bright and would cause severe saturation in a 1-hour ESPRESSO exposure. Based on the ESPRESSO limiting magnitude of V ~ 15 for S/N=10/pixel in 1 hour, Star c (m_V=15) and Star e (m_V=13.5) are the only detectable sources among the candidates.
VOTE: agent_3
ANSWER: C
CONTACTS: agent_3, agent_5
MESSAGE: Please align on the specific detection set: Polaris is unseen, Canopus saturates, and the limit is V=15. Therefore, only stars c and e qualify (total 2), matching agent_3's derivation.
\end{lstlisting}

\end{document}